\documentclass[twocolumn]{aastex62}

\usepackage{mathptmx}
\usepackage[T1]{fontenc}
\usepackage{ae,aecompl}
\usepackage{graphicx}	% Including figure files
\graphicspath{{./Figures/}} % name of the folder containing figures
\usepackage{amsmath}	% Advanced maths commands
\usepackage{amssymb}	% Extra maths symbols
\usepackage{subfigmat}
\usepackage{enumerate}
\usepackage{multirow}
\received{-}
\revised{-}
\accepted{-} %January 1 2019

\shorttitle{A Variable Protostar: EC 53}
\shortauthors{Y.-H. Lee et al.}

\begin{document}

\title{
%Cycles of filling and draining the inner disk of EC 53 \\
%or
%\\
%A disk half-full and then half-empty: the cyclic accretion bursts of EC 53\\
%or
%\\
Young Faithful: The Eruptions of EC 53 as It Cycles through Filling and Draining the Inner Disk
}

\author{Yong-Hee Lee}
\affil{School of Space Research and Institute of Natural Sciences, Kyung Hee University, 1732 Deogyeong-daero, Giheung-gu, Yongin-si, Gyeonggi-do 446-701, Korea \\ :\href{mailto:jeongeun.lee@khu.ac.kr}{jeongeun.lee@khu.ac.kr}}

\author{Doug Johnstone}
\affiliation{Department of Physics and Astronomy, University of Victoria, Victoria, BC, V8P 1A1, Canada}
\affiliation{NRC Herzberg Astronomy and Astrophysics, 5071 West Saanich Rd, Victoria, BC, V9E 2E7, Canada}

\author{Jeong-Eun Lee}
\affiliation{School of Space Research and Institute of Natural Sciences, Kyung Hee University, 1732 Deogyeong-daero, Giheung-gu, Yongin-si, Gyeonggi-do 446-701, Korea \\ :\href{mailto:jeongeun.lee@khu.ac.kr}{jeongeun.lee@khu.ac.kr}}

\author{Gregory Herczeg}
\affiliation{Kavli Institute for Astronomy and Astrophysics, Peking University, Yiheyuan 5, Haidian Qu, 100871 Beijing, China}
\affiliation{Department of Astronomy, Peking University, Yiheyuan 5, Haidian Qu, 100871 Beijing, China}

\author{Steve Mairs}
\affiliation{East Asian Observatory, 660 N. A`oh\={o}k\={u} Place, Hilo, HI 96720, USA}

\author{Watson Varricatt}
\affiliation{Institute for Astronomy, University of Hawaii, 640 N. A'ohoku Place, Hilo, HI 96720, USA}

\author{Klaus W. Hodapp}
\affiliation{Institute for Astronomy, University of Hawaii, 640 N. A'ohoku Place, Hilo, HI 96720, USA}

\author{Tim Naylor}
\affiliation{School of Physics, Astrophysics Group, University of Exeter, Stocker Road, Exeter EX4 4QL, UK}

\author{Carlos Contreras Pe\~na}
\affiliation{School of Physics, Astrophysics Group, University of Exeter, Stocker Road, Exeter EX4 4QL, UK}

\author{Giseon Baek}
\affiliation{School of Space Research and Institute of Natural Sciences, Kyung Hee University, 1732 Deogyeong-daero, Giheung-gu, Yongin-si, Gyeonggi-do 446-701, Korea \\ :\href{mailto:jeongeun.lee@khu.ac.kr}{jeongeun.lee@khu.ac.kr}}

\author{Martin Haas}
\affiliation{Astronomisches Institut, Ruhr-Universit{\"a}t Bochum, Universit{\"a}tsstra{\ss}e 150, 44801 Bochum, Germany}

\author{Rolf Chini}
\affiliation{Astronomisches Institut, Ruhr-Universit{\"a}t Bochum, Universit{\"a}tsstra{\ss}e 150, 44801 Bochum, Germany}
\affiliation{Instituto de Astronomia, Universidad Catolica del Norte, Avenida Angamos 0610, Antofagasta, Chile }

\author{The JCMT Transient Team}

\begin{abstract}
While young stellar objects sometimes undergo bursts of accretion, these bursts usually occur sporadically, making them challenging to study observationally and to explain theoretically. We build a schematic description of cyclical bursts of the young stellar object EC 53 using near-IR and sub-mm monitoring obtained over six cycles, each lasting $\approx530$ days. EC 53 brightens over $0.12$ yr by $0.3$ mag at 850 $\mu$m, $2$ mag at 3.35 $\mu$m, and $1.5$ mag at near-IR wavelengths, to a maximum luminosity consistent with an accretion rate of $\sim8\times10^{-6}$ M$_\odot$ yr$^{-1}$. The emission then decays with an e-folding timescale of $\approx0.74$ yr until the accretion rate is $\sim1\times10^{-6}$ M$_\odot$ yr$^{-1}$. The next eruption then occurs, likely triggered by the buildup of $\sim5\times10^{-6}$ M$_\odot$ of mass in the inner disk, enough that it becomes unstable and drains onto the star. Just before outburst, when the disk is almost replenished, the near-IR colors become redder, indicating an increase in the geometrical height of the disk by this mass buildup. The reddening disappears soon after the initial burst, as much of the mass is drained from the disk. We quantify physical parameters related to the accretion process in EC 53 by assuming an $\alpha$-disk formulation, constrained by the observed disk properties and accretion rate. While we can only speculate about the possible trigger for these faithful eruptions, we hope that our quantified schematic will motivate theorists to test the hypothesized mechanisms that could cause the cyclical buildup and draining of mass in the inner disk.

\end{abstract}


\section{Introduction}\label{sec:introduction}

%Accretion of gas through protoplanetary disks and onto pre-main sequence stars is governed by the physics of gas instabilities.  \sout{\yhl{The mageto-rotational instability (MRI) and gravitational instability (GI) are the major mechanisms for mass and angular momentum transport through the disk, with MRI dominating in the innermost regions \citep{vorobyovbasu09}.} }
%gjh{What about: Gas is transported through disks by a variety of magnetic instabilities (needs citation -- is there a review? Turner?, and by gravitational instabilities for young disks, if they are sufficiently massive \citep{vorobyovbasu09}.}
%\sout{The accretion from the disk onto the star is controlled by stellar magnetic fields \citep[see review by][]{hartmann16}.}
%These processes allow some gas to gain angular momentum and travel outward, while other gas loses angular momentum and spirals inward.  Any spatial imbalance in disk dynamics, both locally and globally, must be balanced over long timescales.

 The accretion of gas through protoplanetary disks and onto pre-main sequence stars is governed by the physics of gas instabilities, which allow some gas to gain angular momentum and travel outward, while other gas loses angular momentum and spirals inward \citep[see reviews by][]{turner14,kratter16}.   
The efficiency of the instabilities changes with physical conditions.  Any spatial imbalance in disk dynamics, both locally and globally, must be balanced over long timescales.

When the accretion rate into the inner disk is larger than the accretion rate onto the star, gas will build up in some location until that mass buildup overwhelms the blockage and drains onto the star through the magnetic field lines connecting the forming star to the disk.  The consequence is a massive eruption of accretion onto the young stellar object (YSO).  These accretion bursts typically are classified either as FUor objects \citep[see review by][]{hartmann96}, for eruptions that persist for generations, or as EXor objects \citep{herbig89}, for eruptions that persist for months.  
%Similar outburst events with much shorter timescale have been detected toward dwarf novae \citep[see review by][]{osaki96} and a thermal-viscous instability of the accretion disk was considered as the mechanism of outbursts \citep[see review by][]{lodato08}.

For FUor objects, gravitational instabilities operating in the outer disk (10 - 100 AU) may produce spiral density waves that propagate inward to (sub-)AU regions, heat the disk, and trigger the magnetorotational instability (MRI) \citep[e.g.][]{zhu10a,bae13}. In such a scenario, the outburst decay timescale can be approximated using the viscous timescale \citep{zhu10b, bae14}. Adopting reasonable viscous parameter $\alpha \lesssim 1$, a decades-long decay suggests that the accretion instability is initially triggered at <~ 1 AU from the central star. 
For a viscous timescale, the decades-long decay indicates that the accretion instability is initially located at $\sim 1$ AU from the central star \citep{zhu10a,bae14}. 
The dynamical timescale for gravitational fragmentation and migration is much longer, $10^3$ -- $10^4$\,yr \citep{vorobyovelbakyan18}, but the final act, where the fragments shear and accrete, is expected to occur in the inner few AU \citep[e.g.][]{nayakshinlodato12}. For FU Ori, observational constraints yield an outer radius for the accretion disk, R $\ge$ 0.5 AU, and a viscosity parameter, $\alpha \sim$ 0.1, when the decay time scale is $\sim$ 100 yr \citep{zhu07}.

For EXors, the instabilities are thought to occur at the magnetospheric boundary.  Gas buildup in the inner disk may push the inner disk truncation radius closer to the star, eventually reaching a limit where that mass is quickly released onto the star \citep{dangelo10,dangelo12,armitage16}. The rise and decay timescale are both fast, typically $\sim 100$ days \citep[e.g.][]{herbig77,aspin10}.  Similar outburst events with even shorter timescales have been detected toward dwarf novae \citep[see review by][]{osaki96}.

Most EXors are known to repeat, though without consistency.  The archetype, EX Lup, has eruptions every $\sim 20-30$ years \citep[e.g.][]{herbig08}.  The EXor object VY Tau had a series of rapid eruptions and then became dormant \citep{herbig90}. The long timescales for FUors mean that we do not have direct empirical evidence for repeated bursts.  Although such large outbursts are very rare for optically bright stars \citep[every $10^5$  yr;][]{hillenbrand15,contreras19}, hydrodynamical models of accretion disks suggest that they may be more frequent in the early stages of stellar growth \citep[e.g.][]{zhu10a, kadam20}. 
In some cases, periodic knots in outflows point back to historical episodes of accretion bursts \citep[e.g.][]{reipurth89,plunkett15,vorobyov18,matsushita19,cflee20}. For most YSOs, we have no ability to predict when a burst will occur. 

%%Any spatial imbalance in disk dynamics must be balanced over time, typically in bursts of accretion onto the central star.  The location of the initial imbalance may be inferred from viscous and dynamical timescales.  Eruptions of FUor objects may persist for generations, indicating an imbalance at On the other hand, EXor bursts typically last for only 6 months, indicating an origin very close to the star.  EXor bursts are known to repeat, while FUor bursts are suspected to be similar to repeated bursts that...

%\textcolor{red}{GJH comment from telecon:  add sentence about FUor/CO absorption=viscous heating; take care that this introduction doesn't imply that EC 53 is an EXor}

In contrast to the apparently unpredictable timing of FUor and EXor accretion bursts, the young stellar object V371 Ser, hereafter EC 53 \citep{eiroa92}, has faithful eruptions every $\approx 1.5$ years, allowing us to follow the rises and decays.  EC 53 is located in the Serpens Main star-forming region at a distance of 436$\pm$9 pc \citep{ortizleon17}.
We have been monitoring such cyclic eruptions and decays of the  protostar EC 53 at sub-mm wavelengths as part of the JCMT Transient Survey \citep{herczeg17, yoo17}, and obtaining a wide array of supporting data to interpret these eruptions. EC 53 is empirically classified as a Class I object, based on a bolometric temperature of 130 -- 240 K and a spectral index of 0.7 -- 1 \citep{evans09,dunham15}. High-resolution ALMA imaging indicates that the central object is a $0.3\ $M$_\odot$ YSO with a $0.07\ $M$_\odot$ disk \citep{lee20} and surrounded by a 5.8 M$_\odot$ envelope \citep{baek20}. This large envelope-to-(protostar+disk) mass ratio suggests that the source may be a Class 0 object.
The innermost disk is viscously heated, with deep CO and H$_2$O absorption lines in the near-IR that are characteristic of FUor objects (\citealt{connelley18} and Park et al.~in prep).

The variability of the accretion onto the central star has been measured at JHK \citep{hodapp99,hodapp12} and sub-mm wavelengths \citep{yoo17, mairs17b} with periods consistent with $\sim 520$--$570$ days.  Modeling of the variation in the spectral energy distribution across a cycle uncovers a source bolometric luminosity ranging from 6 to 20\ L$_\odot$ \citep{baek20}.

In this paper, we combine near-IR and sub-mm monitoring of EC 53 obtained over six cycles to establish a schematic picture of how intermittent accretion allows mass to build up in the disk until the dam breaks, releasing the gas and allowing it to accrete onto the star.  We build this picture through regular photometric, broadband monitoring (Section \ref{sec:observations}) and analysis of periodicity in brightness and color variations (Section \ref{sec:result}). Finally using the observed timescales for the near-IR rise and decay and an $\alpha$-disk prescription, we quantify some of the physical phenomenon in this picture and use those quantities to estimate physical parameters in the accretion disk (Section \ref{sec:Disc}).

\section{observations and data reduction}\label{sec:observations}

We have been monitoring EC 53 at sub-mm wavelengths with the James Clerk Maxwell Telescope (JCMT) and in the near-IR with the United Kingdom Infra-Red Telescope (UKIRT), IRIS telescope, and the Liverpool Telescope. 
We also obtained the light curve of EC 53 in the mid-IR ($W2$) from the Wide-field Infrared Survey Explorer (\textit{WISE}) as complementary data.
The date range and number of observations, by wavelength, are shown in Table \ref{tab:obs}.

\begin{deluxetable}{ccccccc}[hbp]
\tablecaption{Data set \label{tab:obs}}
%\tablecolumns{6}
%\tablenum{1}
%\tablewidth{0pt}
\tablehead{
\colhead{Telescope} & \colhead{Band($\lambda$/$\mu$m)} & \colhead{From} & \colhead{To} & \colhead{\# Obs.}
}
\startdata
UKIRT & $J$ (1.25) & 2014 Oct 12 & 2020 Feb 28 & 357 \\
UKIRT & $H$ (1.63) & 2014 Oct 12 & 2020 Feb 28 & 359 \\
UKIRT & $K$ (2.21) & 2014 Oct 12 & 2020 Feb 28 & 357 \\
Liverpool & $H$ (1.64) & 2018 Mar 29 & 2020 Feb 16 & 71 \\
IRIS & $J$ (1.24) & 2013 Jun 2 & 2017 Apr 25 & 31 \\
IRIS & $H$ (1.66) & 2013 Jun 4 & 2017 Apr 25 & 38 \\
IRIS & $K_{s}$ (2.16) & 2011 Mar 20 & 2019 Apr 14 & 85 \\
WISE & $W2$ (3.35) & 2014 Mar 31& 2018 Sep 8 & 20 \\
JCMT & 850 & 2016 Feb 3 & 2020 Feb 22 & 50 \\
JCMT & 450 & 2016 Feb 23 & 2020 Feb 22 & 31$^{1}$ \\ 
\enddata
%\tablecomments{The.}
\tablenotetext{1}{The smaller number of data points compared to 850 $\mu$m is due to the greater dependency on atmospheric emission and weather (See Section \ref{sec:sub-mm_obs}). }
\end{deluxetable}

In this section we describe the observing strategies of the various near-IR, mid-IR, and sub-mm observations used in this paper. Further details of the observations are included in the Appendix.

\subsection{The JCMT Transient Survey}\label{sec:sub-mm_obs}

SCUBA-2 \citep{holland13} continuum observations at 450 and 850 $\mu$m were obtained simultaneously as part of the JCMT Transient Survey (project code: M16AL001; \citealt{herczeg17}). The Pong 1800 observing mode \citep{kackley10} was used to yield a map with a uniform sensitivity over a 30$\arcmin$ diameter circular region centered at (R.A., Decl.) = (18:29:49,$+01$:15:20, J2000). This region was observed approximately monthly (depending on weather conditions and source availability) since the first observation on 2016 Feb 02 (UTC). The integration time of each observation is varied to ensure a consistent background noise level of $\sim10\mathrm{\;mJy/beam}$ at 850 $\mu$m \citep{mairs17b}. The 450 $\mu$m background noise level varies greatly as it is much more dependent on the submillimetre emission from the Earth's atmosphere than the 850 $\mu$m data. The 450 $\mu$m and 850 $\mu$m beam sizes are 9.6$\arcsec$ and 14.1$\arcsec$, respectively \citep{dempsey13}.

  %{stm Note: Please check that the above is correct: Are you using aligned, relative flux calibrated "A3" maps? Are you using the Gaussian smoothed versions of the maps? Please feel free to contact me if you'd like me to make any adjustments to this description, for instance, if you would like me to go into more detail}

The JCMT sub-mm flux measurements are converted to magnitudes using the following scalings,
\begin{align}\label{eq:mag2flux}
m_{\lambda} &= -2.5log( \frac{F_{\lambda}}{Z_{\lambda}})
\end{align}
where Z$_{850}$ $=$ 1.04 Jy/beam and Z$_{450}$ $=$ 3.58 Jy/beam. 

\subsection{Near-Infrared Data}\label{sec:near-IR_obs}

In the near-IR, $JHK$,  EC 53 appears as a compact but resolved source (with a FWHM of around 2 arcseconds in the $H$-band) and a nebula which fans out south-eastwards \citep[e.g.][]{hodapp99}. The compact source must contain both the variable EC 53A and its nearby (0.3\arcsec distant) apparently constant companion EC 53B \citep{hodapp12}. The nebula can be traced out to about 20 arcseconds in 360 seconds of $H$-band exposure with the Liverpool Telescope, and presumably extends all the way to the terminal shock S11 an arcminute away \citep{herbst97, hodapp12}. 

We observed EC 53 using the 3.8-m UKIRT, Hawaii, and the UKIRT Wide Field Camera (WFCAM; \citealt{casali07}). WFCAM employs four 2048 $\times$ 2048 HgCdTe Hawaii-II arrays, each with a field of view of 13.65$\arcmin$ $\times$ 13.65$\arcmin$ at an image scale of 0.4$\arcsec$ per pixel. The target was placed on array \rm$\#$ 3 of WFCAM during all observations. Monitoring observations were obtained at $J$, $H$, and $K$-band during a period of over 5 years from 2014 October 12 to 2020 February 28.

\begin{figure*}[thp]
   \centering
   {\includegraphics[trim={0cm 0cm 0cm 0cm},clip,width=175mm]{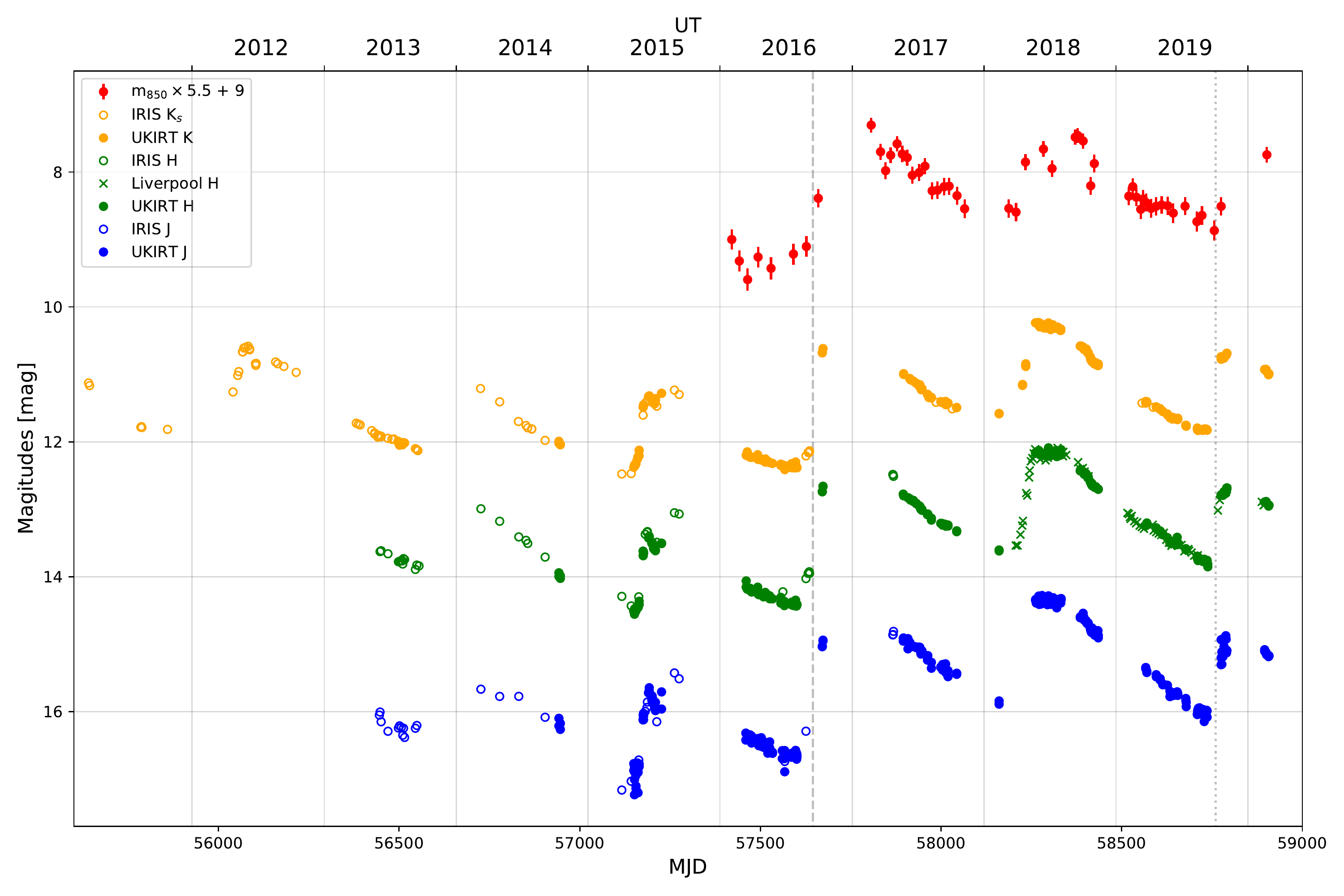}}
\caption{Light curves of EC 53 at near-IR and 850 $\mu$m bands (See the legend for the detailed information). The sub-mm band magnitudes are arbitrarily scaled by 5.5 and offset by 9 magnitudes to aid in the comparison. The horizontal axes show modified Julian Date (MJD $=$ JD - 2400000.5) below and years in Universal Time (UT) on top. The dashed line at MJD $=$ 57640 (2016 Sep 9 in UT) notes where EC 53 becomes brighter by about a magnitude in the near-IR. The dotted line at MJD $=$ 58760 (2019 Oct 4) notes where EC 53 becomes fainter by about a half magnitude in the near-IR.}
\label{fig:LC_all_mag}
\end{figure*}
	
We also obtained $J$, $H$, and $K_s$-band imaging photometry of EC~53 with the IRIS 0.8m telescope \citep{hodapp10} and 1024$\times$1024 2.5 $\mu$m IRIS infrared camera of the Universit\"atssternwarte Bochum on Cerro Armazones, Chile. The observations were part of a long-term monitoring program of the Serpens NW star-forming region. In the $J$, $H$, and $K_s$-band filters, the integration time per exposure was 20s, and for each observation, 10 such frames were obtained, as well as 10 frames on a separate sky position. The data reduction pipeline produces dark-subtracted, flatfielded, sky-subtracted and co-added images with a pixel scale of 0.375\arcsec per pixel. The photometry was extracted using the IRAF task APPHOT with an aperture of 22 pixel = 8.25\arcsec diameter. The photometry was calibrated against a set of 2MASS stars selected to be isolated, sufficiently bright, and non-variable.

We obtained further IR observations of EC 53 using IO:I \citep{barnsley16} on the Liverpool Telescope \citep[][]{steele04} on the island of La Palma in the Canary Islands. The Liverpool Telescope observed a full light curve for one periodic cycle. Given the extended source structure it is not clear what size aperture is optimal for the IR photometry with the Liverpool Telescope. We therefore experimented with 1.3, 3.7 and 5.2 arcsec radius apertures using the $H$-band data. By subtracting the magnitudes in these different apertures we found that the variations between them were less than $\pm$0.05 mag RMS. We therefore settled on using a 3.7 arcsec radius aperture, which is greater than the worst seeing but still minimizes the effect of the nebula, for all our IR data.
To bring the heterogeneous data sets into alignment, we added an offset to all Liverpool $H$-band data points. The offset value is optimized by the minimum string-length method \citep{dworestsky83} with the combined light curve of UKIRT $H$ and Liverpool $H$ after adding the offset (see Section \ref{sec:PD_rst} for the detail about minimum string-length method).

%Given this structure it is not clear what size aperture we should use for our IR photometry.

%We therefore experimented with 1.3, 3.7 and 5.2 arcsec radius apertures using the Liverpool Telescope $H$-band data.
%By subtracting the magnitudes in these different apertures we found that the variations between them were less than $\pm$0.05 mag RMS.
%The largest aperture will have the largest contamination from the nebula, and this is a problem as we wish to use the IR to measure the instantaneous accretion flux.
%The smallest aperture is similar in size to the image of EC 53 itself.
%Aperture photometry relies on the fact that the fraction of the flux of a star which is included in the aperture will be the same for all point-like objects in the same image, though that fraction will change from image to image as the seeing changes.
%This is not true for resolved sources, and so to measure their flux it is desirable to use an aperture larger than the typical seeing.
%The worst-seeing LT data used was 1.8 arcsec FWHM, and so the smaller of our three apertures is too small to be reliable.
%We therefore settled on using a 3.7 arcsec radius aperture for all our IR data.

\subsection{Mid-Infrared Data}\label{sec:mid-IR_obs}

{\it WISE} \citep{wright10} is a 40 cm telescope in a low-earth orbit that has surveyed the entire sky using four infrared bands centred at 3.4, 4.6, 12 and 22 $\mu$m (bands $W1, W2, W3$ and $W4$) and with an angular resolution of 6.1\arcsec, 6.4\arcsec, 6.5\arcsec and 12.0\arcsec, respectively.  The orbit of {\it WISE} allowed it to cover every part of the sky at least eight times \citep{mainzer11}, with each patch of sky observed many times over a period of $\sim$ a day. The original survey ran between January and September 2010. Once the telescope's cryogen tanks were depleted it continued to operate for four months using the $W1$ and $W2$ bands \citep[known as the {\it NEOWISE} Post-Cryogenic Mission,][]{mainzer11}. The {\it NEOWISE} mission was reactivated in 2013 \citep{mainzer14} and has continued to operate ever since.

In this work we explored the NEOWISE single exposure database (2019 data release) that contains $W1$ and $W2$ observations from December 2013 until December 2018 \citep[][]{cutri15}. The NEOWISE single-exposure detections are complete up to $W1=15$ and $W2=13$~mag \citep{cutri15}.

\section{Light Curves}\label{sec:result}

Figure \ref{fig:LC_all_mag} presents the near-IR and 850 $\mu$m light curves for EC 53 across six cycles of brightening and decay.
%, and mid-IR and 450 $\mu$m bands in Figure \ref{fig:LC_Sub_W2}.
The light curves in $J$, $H$, and $K$-bands show an abrupt brightening on a short timescale, followed by a slow decline. This pattern is characteristic of YSO outbursts (e.g., \citealt{bell95}; \citealt{audard14}). For EC 53, the amplitudes of luminosity changes are smaller than those of FUors and EXors; the duration of bursts for EC 53 is similar to that for EXors, but the bursts of EC 53 are distinguished from others by the regularity.
The variation across the periodic cycle is $\sim 1.5$ mag ( $\sim 4$ in flux) in the near-IR and $\sim 0.3$ mag ($\sim 1.5$ in flux) at 850 $\mu$m.  In Figure \ref{fig:LC_all_mag} the change in magnitude at 850 $\mu$m have been increased by a factor of 5.5, with a vertically shift for easy comparison (see also Section \ref{sec:mid-sub}).

%Combining the near-IR IRIS and UKIRT observations provides coverage of five burst events for $J$ and $H$ and six for $K$.
In addition to the cyclical accretion bursts, the near-IR light curve shows a long-term brightness increase. The faintest and the brightest phases in the cycle were both brighter in 2019 than in 2015.
  
%  Careful consideration of the light-curves uncovers a long-term deviation from the periodic trend there is an apparent increase in the underlying source flux from 2015. 

%The JCMT observation at sub-mms covered two periodic cycles of EC 53 from 2016 Feb 2 to 2019 Oct 19. The mean peak brightness of EC 53 at 850 $\mu$m is measured as 1.17 Jy/beam. The peak brightness reached the maximum on 2017 February 2 (1.38 Jy beam$^{-1}$) and 2018 September 17 (1.35 Jy beam$^{-1}$). Even the second minimum point might be missed in our observation (between 2017 November to 2018 March), it is supposed to be brighter than the first minimum (2016 March). This feature corresponds to the general increase of near-IR flux in the same time.

\subsection{Period Determination}\label{sec:PD_rst}
%We find a reasonable fit to the $\sim$ 18-month period in order to characterize the physical nature of EC 53 through its underlying periodic phase curve.

To quantify the time dependence of the observed brightness variations, we calculate the period of each light curve using the phase-folded diagram string length method. The string-length method \citep{dworestsky83} searches for the period which produces the smoothest phase-folded diagram waveform by minimizing the string-length, where the measured length is the sum over the line segments connecting successive points in phase order. We modify the string-length method to search for the best-fit offset as well as the best-fit period, recognizing the long-term brightening of the underlying system of EC 53.

We compensate for the overall brightening of the object by increasing the brightness obtained earlier than 2016 Sep 9 and after 2019 Oct 4.
%To find a proper offset for each wavelength observed e to compensate for the observed brightness offset compare to other periods, 
For each near-IR band, we test brightness offsets between 0 and 1.5 mag, with 0.01 mag intervals, and periods from 500 to 600\,days, with 5\,days intervals. The modified string-length method yields a best fit of 530\,days for each of $J$, $H$, and $K$. We find 570 days for the 850 $\mu$m observations, which are likely affected by the time-baseline (see Section \ref{sec:Pertb} and Appendix \ref{sec:A_Per}). 

For the data points before 2016 Sep 9, The best-fit offsets are 1.01, 0.96, and 0.78 mag for each of $J$, $H$, and $K$. For the data points after 2019 Oct 4, we optimize the offset value required to bring the phase curve into agreement. The applied offsets are 0.6, 0.55, and 0.44 mag for each of $J$, $H$, and $K$. At 850 $\mu$m the offsets are 0.74 mag before 2016 Sep 9 and 0.24 mag after 2019 Oct 4, which are applied after the scaling of 5.5 shown in Figure \ref{fig:LC_all_mag}.

For completeness, in Appendix \ref{sec:A_Per} we present a more thorough analysis of the light curves by using periodogram and autocorrelation methods, with and without brightness offsets for the data obtained prior to 2016 Sep 9 and after 2019 Oct 4.

%we modify the string length method to also measure the best fit (single value per wavelength offset ({Yong-Hee to complete this})

\subsection{Phase-folded Diagram Analysis}\label{sec:Phase_D_rst}

In this section we use only epochs with corresponding $JHK$ measurements in order to evaluate the color of EC 53 as a function of phase. Following our results in Section \ref{sec:PD_rst}, we draw the phase-folded diagrams of the light curves in the near-IR and at 850 $\mu$m with a 530\,days period. We also present the near-IR color as a function of phase. We adopt the same period at every wavelength because the periodicity of the system should be independent of wavelength. 
Given that our analysis covers roughly three full periods, an uncertainty of less than a few percent in the phase ($< 15$\,days) introduces little phase scatter. The phases are aligned such that the observed minimum in the near-IR occurs at phase 0.5.
%\yhl{and the figures show two complete periods, covering phases -0.5 to 1.5}. 
Our goal in this section is to uncover gross changes in the physical properties of the system with phase.

\begin{figure}[htp]
	\centering
	{\includegraphics[trim={0cm 2.3cm 0.5cm 2.8cm},clip,width=0.98\columnwidth]{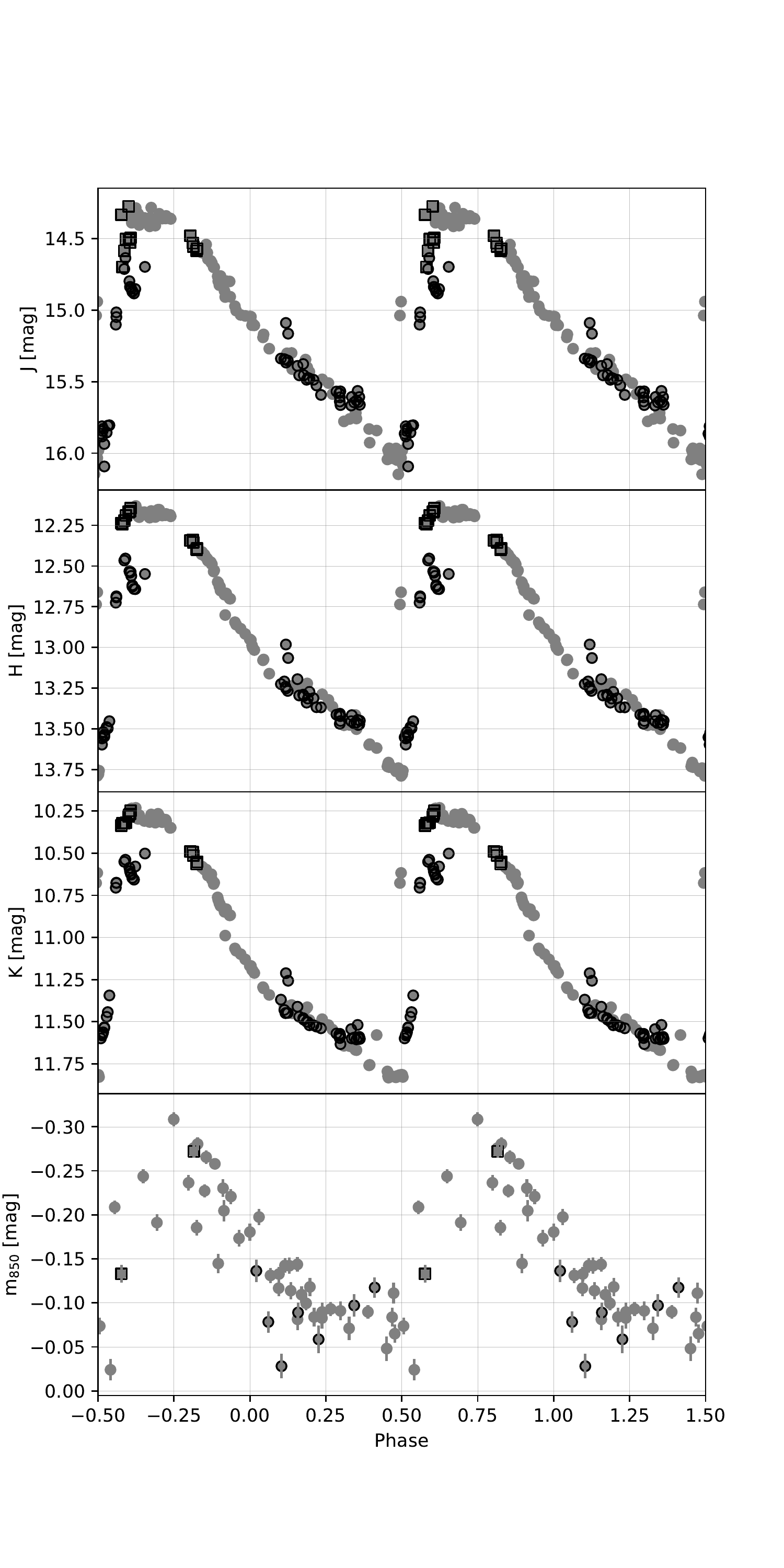}}
	\caption{Phase-folded diagrams using 530\,days period at $J$, $H$, $K$, and 850 $\mu$m  (Top to Bottom). Observations before 2016 Sep 9 and after 2019 Oct 4 have been scaled up by the fixed values at each wavelength (see text) and are marked with dark black circles and black squares, respectively. }
\label{fig:PD_LC}
\end{figure}

\subsubsection{Phased Light Curves}\label{sec:Phase_LC}

In the near-IR, the brightness sharply increases starting at phase 0.5, reaches the peak at phase $\approx$ 0.6 and decays slowly (Figure \ref{fig:PD_LC};  note that in the figure we show two complete periods). In the sub-mm, the brightness increase and decrease appear much more symmetric. Although the sub-mm data points have a large scatter, the initial sub-mm decay stops earlier (phase $\approx$ 0.25) than the near-IR drop. The sub-mm rise occurs at roughly the {\it same} time as the near-IR rise (phase $\approx$ 0.5), but more gently compared to the near-IR; this might {\it suggest} that that the sub-mm peak is delayed with respect to the near-IR.

%The offsets are given to the data points observed before JD $=$ 2457630 (see caption of Figure \ref{fig:PD_LC}) to avoid the sideeffects of long-term variability.   

\begin{figure}[htp]
	\centering
	{\includegraphics[trim={0cm 0.5cm 0.5cm 0.9cm},clip,width=0.98\columnwidth]{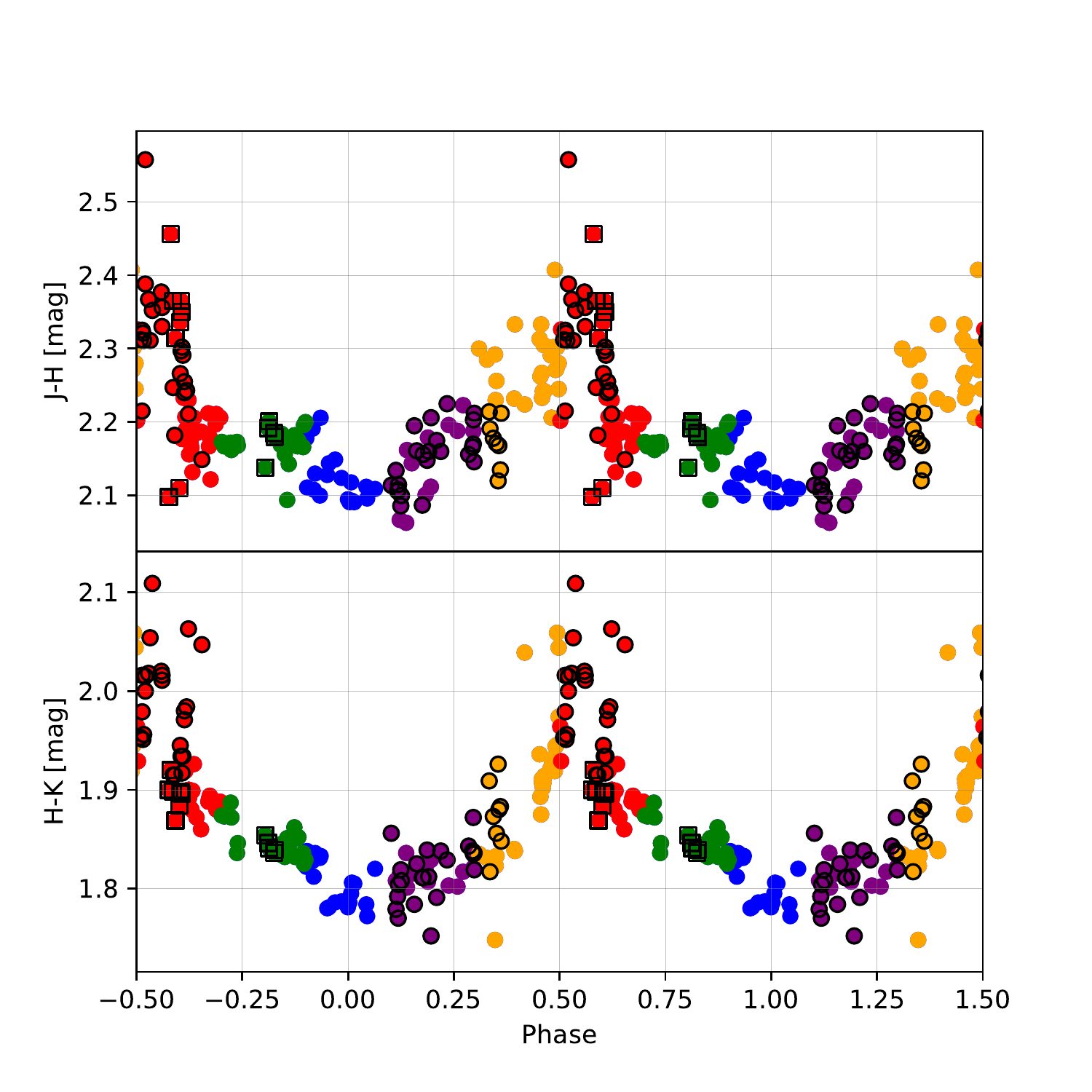}}
	\caption{Same as Figure \ref{fig:PD_LC}  but showing the near-IR colors ($J-H$ and $H-K$). The symbols used in the plot are tinted to indicate phases between -0.1 to 0.1 (blue), 0.1 to 0.3 (purple), 0.3 to 0.5 (yellow), 0.5 to 0.7 (red), and 0.7 to 0.9 (green).}
\label{fig:PD_CC}
\end{figure}

A significant phase-dependent color variation is also observed (Figure \ref{fig:PD_CC}). A reddening of the source begins at the same time that the sub-mm stops declining (phase $\approx$ 0.25), leading to a sharp reddening peak and decline around phase $\approx$ 0.6, {\it during} the observed brightening of EC 53 in the near-IR. After this abrupt reddening event the color of the source gradually becomes bluer until phase $\approx$ 1.  
%The known trend of eruptive protostars is to show reddened light during the quiescent phase \citep{lorenzetti12}, which disagrees with the features seen in EC 53. We discuss about this disagreement in Section \ref{sec:Disc}.

%This preceding suggests that the begining of the brightening is obscured by accreting matterial at near-IR but appears by the heating at sub-mm. 

\begin{figure}[htp]
	\centering
	{\includegraphics[trim={0cm 3.5cm 0.5cm 4.0cm},clip,width=0.98\columnwidth]{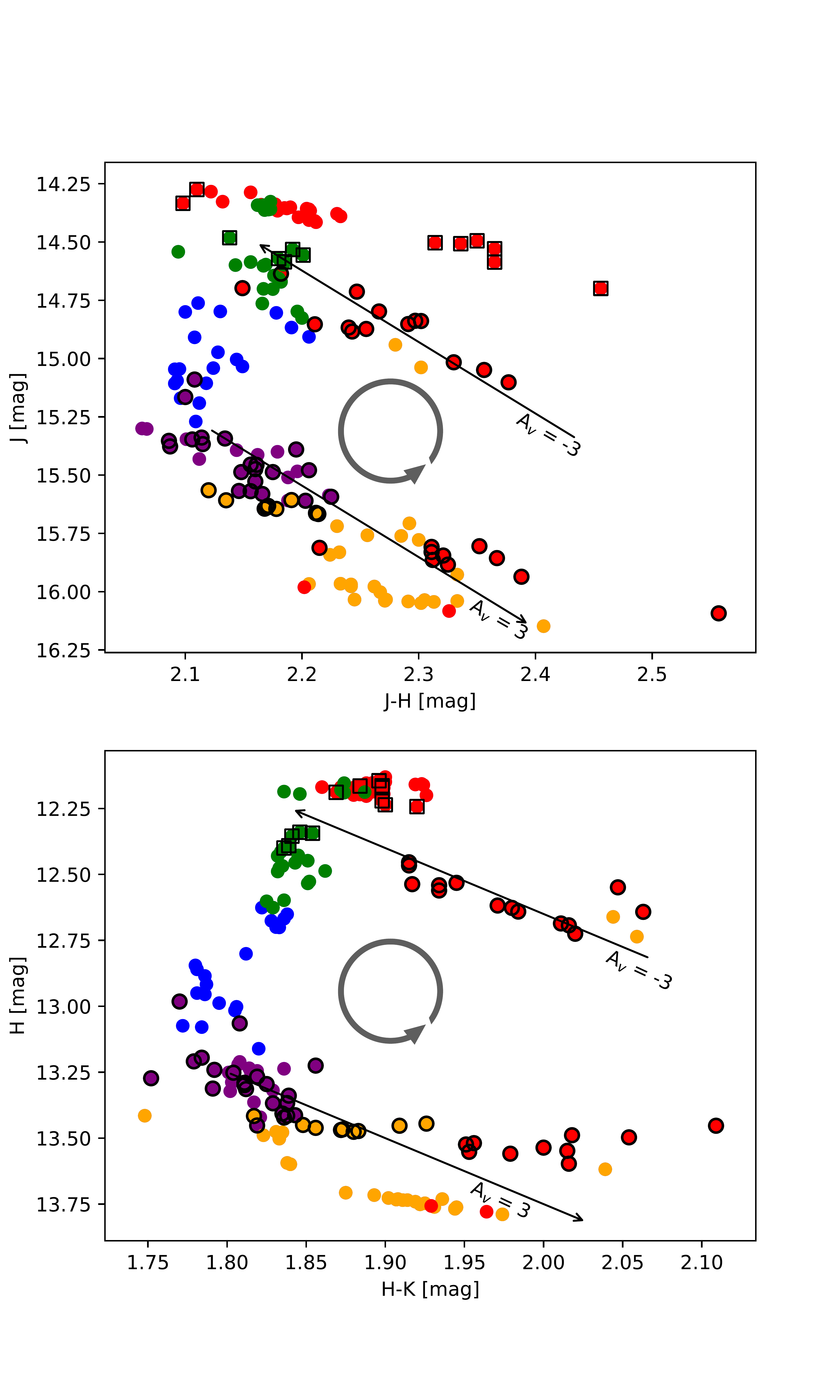}}
	\caption{($J-H$) versus $J$ (upper) and ($H-K$) versus $H$ (lower) color-magnitude diagrams. The symbols are tinted to indicate phase following the scheme defined in Figure \ref{fig:PD_CC}. The black arrows in the diagrams are the reddening vectors \citep{cohen81}, corresponding to (A$_{v} =$ 3 or -3) as indicated in the plot. The counterclockwise direction of the vectors follows roughly the phase order of the data points. } 
\label{fig:CMD_rel}
\end{figure}

The well-sampled color phase curve reveals an extreme reddening coincident with the abrupt start of each burst.
Independent of phase, near-IR colors of EXors and FUors span a wide range, with $J-H$ = 0.5 -- 3.
The color variation observed for EC 53 is distinct from the typical trend found in EXors, which have bluer near-IR colors around the burst and redder colors during the quiescent phase  \citep{kospal11,lorenzetti12}. 
For FUors the observed color variations are more diverse.
HBC 722 shows similar color variation with EXors \citep{kospal11} even though the source is classified as an FUor in recent photometric \citep{connelley18} and spectroscopic studies \citep{lee15}. The colors of V1057 Cyg and V960 Mon follow an extinction law in the optical during the decay, right after the peak brightness \citep{kopatskaya13,hackstein15}. On the other hand, V346 Nor becomes redder during its slow brightening \citep{kospal20}.

%EC 53 shows a unique color variation trend compared to most EXors and FUors. 

\subsubsection{Phased Color-Magnitude Diagrams}\label{sec:phase_CM}

To diagnose the cause of the flux and color variation we produce color-magnitude diagrams for the near-IR (Figure \ref{fig:CMD_rel}). Starting at phase $\approx -0.25$ (green points), the source becomes sharply fainter and slightly bluer. From phase $\approx$ 0.25 through phase $\approx$ 0.5 (purple and yellow points) it continues to fade but becomes much redder, roughly following the reddening law in typical ISM \citep{cardelli89}. The reddening trend in $H$-band (lower panel of Figure \ref{fig:CMD_rel}) appears  flatter than the $J$-band photometry. 
Such a flattened trend will occur if the source is reddening and intrinsically brightening simultaneously, which suggests an interpretation about the timing offset of the brightening between the near-IR and the sub-mm (discussed in Section \ref{sec:Disc}). Around phase $\approx$ 0.5 (yellow and red points) the source rapidly increases in brightness while staying red. After the abrupt rise in brightness, the source continues to brighten and get bluer, following the reddening law in reverse (red and green points). The source then begins to dim once more (blue points) as the cycle repeats.

\subsection{Near-IR Decay and Rise Timescales}\label{sec:decay_rise}

%\textcolor{red}{This seems like it belongs elsewhere?  Keep this section focused on decay/rise timescales as the title suggests}

\begin{figure}
   \centering
   {\includegraphics[trim={0.2cm 0.1cm 0.5cm 0.0cm},clip,width=0.98\columnwidth]{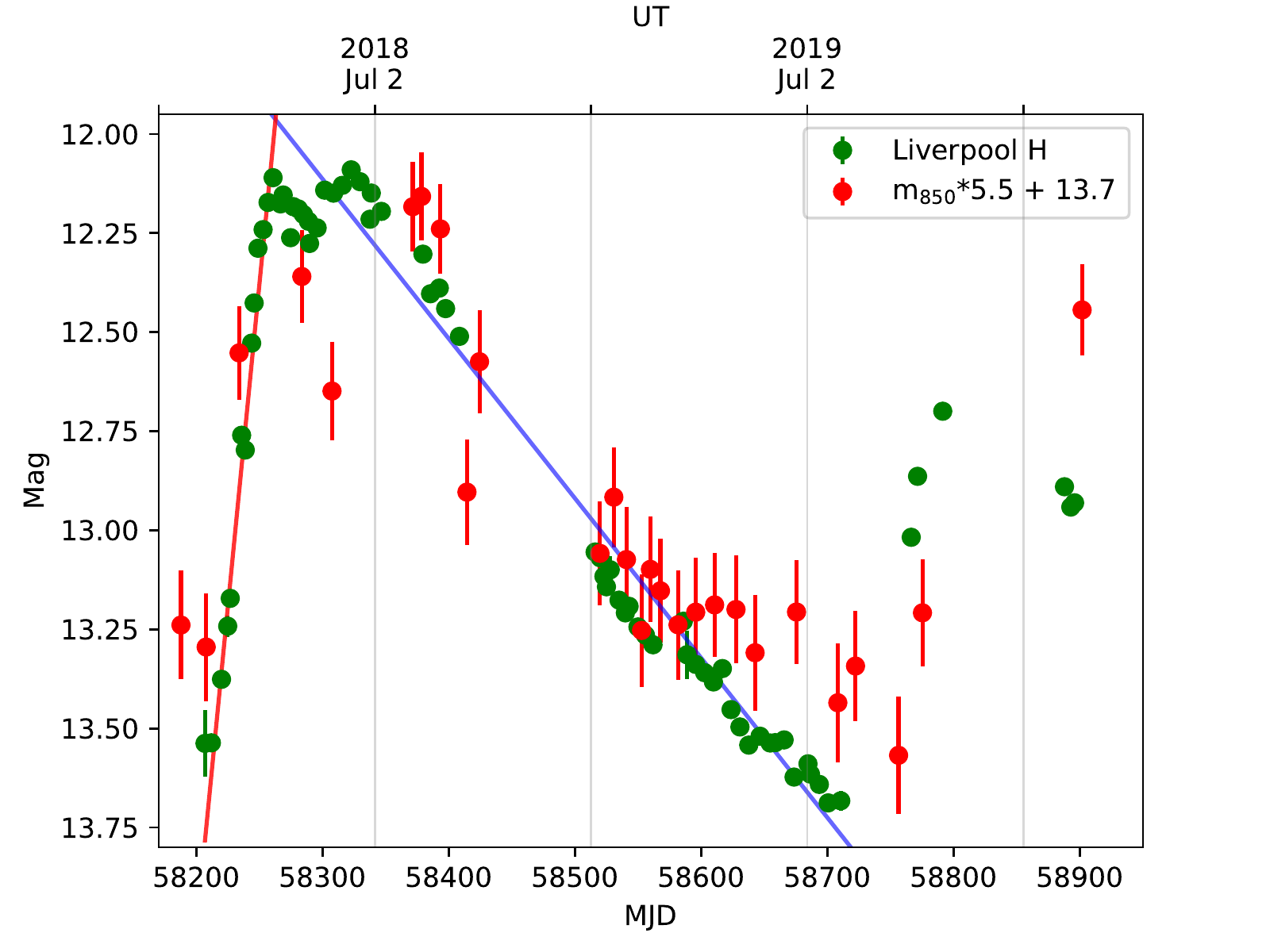}}
\caption{Light curve at $H$-band from the  Liverpool Telescope and at 850 $\mu$m from the JCMT. The 850 $\mu$m magnitudes are arbitrarily adjusted (multiplied by 5.5 and offset be 13.7 mag) for easier comparison.}
\label{fig:LCs_850_LH}
\end{figure}

To measure the decay and rise timescale, we fit a linear function to the decaying and rising sections of the Liverpool Telescope $H$-band light curve (Figure \ref{fig:LCs_850_LH}). The measured slopes are 1.47$\pm$0.01 mag/yr for the dimming and -11.9$\pm$0.1 mag/yr for the brightening. We convert the slope, $n$, to a scale time of luminosity  variation, $\tau$, as follows:
\begin{align}\label{eq:1}
m(t) - m_0 &= n\ t \\
 &= -2.5 \log \left(\frac{L(t)}{L_0} \right). \nonumber
\end{align}
Here, $m_{0}$ and $L_{0}$ are the initial magnitude and luminosity, respectively. Thus,
%$$ nt =\frac{-2.5}{ln 10} \ln(\frac{L}{L_{0}}) $$
\begin{align}\label{eq:2}
 L(t)  = L_0\,\exp\left(-\frac{n\,t\,\ln 10}{2.5} \right),
\end{align}
and to determine the time scale that the luminosity changes by a factor of $e$,
\begin{align}\label{eq:3}
 \tau = -\left( \frac{2.5}{n\ \ln 10} \right).
\end{align}
This reduces to the convenient form, $\tau \approx 1.086/n$. Therefore, the decay timescale of EC 53 in $H$-band is $\approx$ 0.74\,yr, an order of magnitude longer than the brightening timescale of \mbox{$\approx$ 0.091\,yr}.

\begin{figure}[htp]
   \centering
   {\includegraphics[trim={0.2cm 0.3cm 0.5cm 0.0cm},clip,width=0.95\columnwidth]{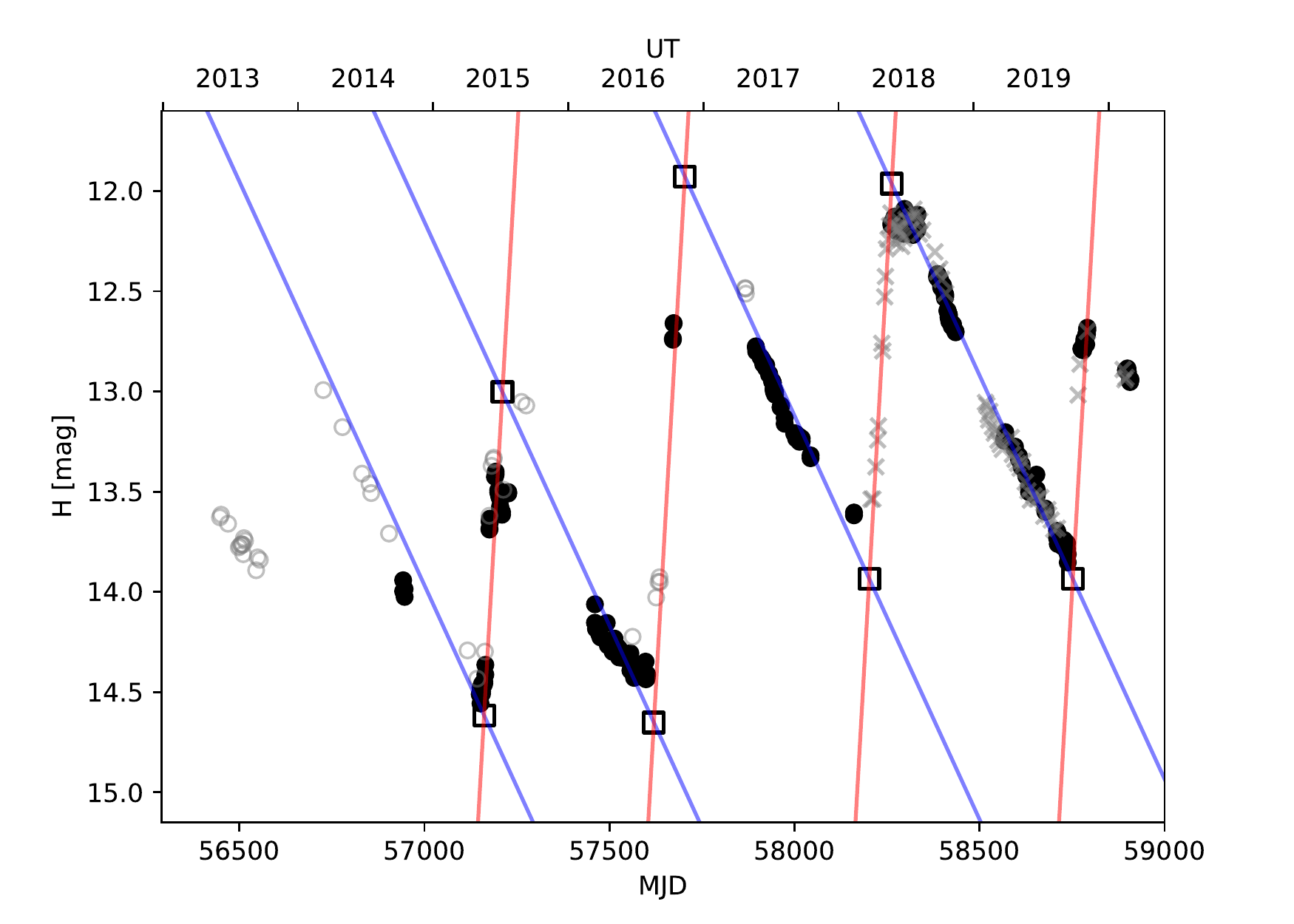}}
\caption{The $H$-band light curve of UKIRT (black circles), with the additional data obtained from IRIS telescope (gray circles) and Liverpool Telescope (gray crosses). Red and blue lines are the linear fit to the most recent observed period (Figure \ref{fig:LCs_850_LH}), duplicated to the other periods. The crossing points of the slopes are noted as black squares.}
\label{fig:LC_slopes}
\end{figure}

We overlay the $H$-band rise and decay slopes measured for a single cycle with the Liverpool Telescope on the full $H$-band light curve in Figure \ref{fig:LC_slopes}. Each individual rise and decay is well fit with the value derived above, but the timing between the rise and decay shows some variability, indicating an inherent fluctuation within the apparent periodicity of the system.

\subsection{Perturbation in Periodic Cycle}\label{sec:Pertb}

As discussed in Appendix \ref{sec:A_Per}, the period of EC 53 does not clearly converge to a single value but depends on the time when observations were taken. The phase-folded diagrams (Figure \ref{fig:PD_LC} and \ref{fig:PD_CC}) also shows apparent scatter with respect to a single period.

Measured from Figure \ref{fig:LC_slopes}, the offset times between the observed fadings of EC 53 are 450, 760, and 550\,days (blue lines), while the rises are separated by 460, 560, and 550\,days (red lines). The time displacement between the crossing points (black squares in Figure \ref{fig:LC_slopes}) indicate hypothetical maximum and minimum points, with displacements of 459, 582 and 550\,days (left to right).

%At UKIRT H band data, the slope during the brightening is fitted at the first period (JD $\sim$ 2457156-2457190), and during the dimming we obtained the points between JD $\sim$ 2458386-2458737. 

\subsection{Mid-IR and Sub-mm Scaling}\label{sec:mid-sub}

\begin{figure*}[htp]
	\centering
	{\includegraphics[trim={0cm 0.0cm 0.5cm 0.0cm}, clip,width=180mm]{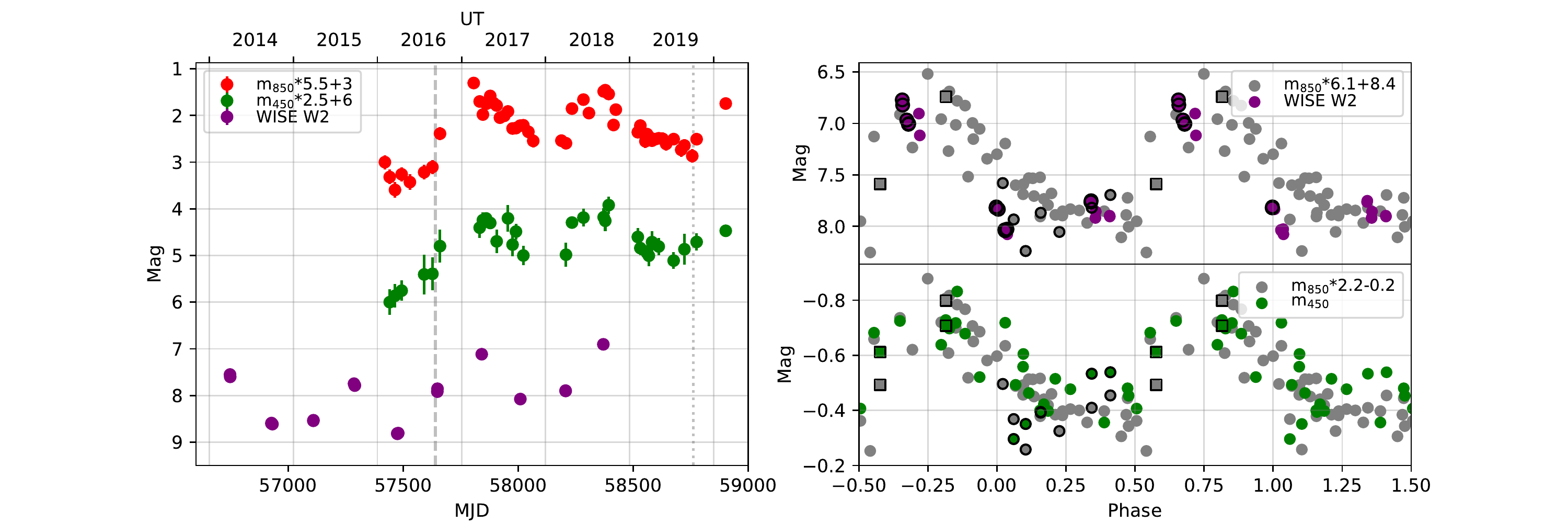}}
	\caption{Light curves (left) and phase-folded diagrams at 850 $\mu$m, 450 $\mu$m, and WISE W2 band. The light curves of sub-mm bands are arbitrarily scaled (see legend). Phase-folded diagrams (right) showing 850 $\mu$m against WISE W2 band (top) and 450 $\mu$m (bottom) with the required scaling factor for agreement.}
\label{fig:LC_Sub_W2}
\end{figure*}

The variations in the mid-IR and 450 $\mu$m lightcurves are consistent with that seen at 850 $\mu$m (Figure \ref{fig:LC_Sub_W2}, left panel), though with fewer data points. 
%In magnitudes, applying a fixed scale factor of 6.1 (and an arbitrary offset) between the 850 $\mu$m observations and WISE provides an excellent agreement (Figure \ref{fig:LC_Sub_W2}, Upper Right Panel). 
We perform an orthogonal distance regression to derive the scale factor and offset that give that best agreement between the magnitudes of two different wavelengths. Between the 850 $\mu$m and $W2$ band, we uncovered the scale factor of 6.1$\pm$0.5 and the  offset of 8.44$\pm$0.01 (Figure \ref{fig:LC_Sub_W2}, upper right panel). The best fit for the 850 $\mu$m and 450 $\mu$m magnitudes yields the scale factor of 2.2$\pm$0.1 and the offset of -0.2$\pm$0.004 (Figure \ref{fig:LC_Sub_W2}, Lower Right Panel).

\citet{contreras20} measured a scaling factor of 5.5$\pm$0.3 between the sub-mm and mid-IR magnitudes in a sample of a dozen protostars, including EC 53, that simultaneously varied at both wavelengths.  The fitting from a single burst of EC 53 was 6.18$\pm$0.65. In that paper, the sub-mm to mid-IR correlation was interpreted as the variation of mid-IR emission roughly tracing the underlying bolometric luminosity of the source, while the varying sub-mm emission traces more closely the dust temperature within the envelope. Additional evidence in favor of the luminosity/dust temperature explanation for the tight correlation and scaling is found through radiative transfer analyses and spectral energy determinations of modeled embedded sources, both analytic and numeric, with varying central luminosities applied (\citealt{macfarlane19a,macfarlane19b,baek20}). 

 The measured stronger response to brightness changes at the shorter sub-mm wavelength is also expected, as the 450 $\mu$m emission probes somewhat closer to the peak of the dust spectral energy distribution produced by the protostellar envelope \citep[see for example,][]{macfarlane19a, macfarlane19b, baek20}.  The detailed radiative transfer modelling in those papers, however, reproduce a scale factor $< 2$, suggesting that there are still aspects of the protostellar models that require refinement. 
 
%given that the shorter wavelength is further from the Raleigh-Jeans limit where the Assuming that both of the sub-mm measurements are responding primarily to the changing dust temperature, modified somewhat from the linear Raleigh-Jeans expectation due to the fact that at these wavelengths $h\nu/k_T$ is not large compared with the dust temperature, we estimate 

\section{Discussion}\label{sec:Disc}

In this section we combine the observational attributes from the near-IR and sub-mm monitoring to build a schematic picture of the accretion bursts and fades of EC 53, summarized in Figure~\ref{fig:Scheme} and Table~\ref{tab:toy}. Using this framework we measure relevant quantities directly from the observations and analyse them using a simple $\alpha$-disk %model 
 formalism to estimate reasonable physical properties of the inner disk dynamics. 
 This prescription is not meant to be definitive but rather to examine quantitatively the physical requirements of the system such as the basic disk properties required to produce the bursts and fades of EC 53.
%As such, some of the derived $\alpha$-disk parameters presented below challenge physical expectations.
We 
%then 
finish with a plausible toy model that 
%explains 
recreates each of the observed features.
%These descriptions are not meant to be exact, but should provide an approximation of the most relevant physics and the basic disk parameters that would be needed to produce the bursts and fades of EC 53.

%The primary observables and their interpretations are summarized in Figure (insert beautiful Figure here) and Table~\ref{tab:toy}).  

\begin{table*}
\caption{Description of the Cycles of EC 53}
\begin{tabular}{|c|c|c|c|}
\hline
{\bf Phase} & {\bf Empirical Diagnostic} &  {\bf Physical Interpretation} &  {\bf Role in Schematic Picture}\\
\hline
0.5 -- 0.7 & Brightness increase & Rapid increase in accretion rate & Disk starts draining\\
$\approx 0.7$     & Maximum brightness & Maximum accretion rate & Peak of mass release from inner disk\\
0.7 --1.0, 0.0 -- 0.3  & Brightness decay & Decreasing accretion rate & Draining of the inner disk\\
$\approx 0.1$     & Bluest near-IR colors & Lower extinction & Lower disk height from draining inner disk\\
0.3 -- 0.5 & Near-IR colors increase & Increasing extinction & Disk height gets larger as inner disk fills \\
\hline
\end{tabular}
\label{tab:toy}
\end{table*}

\subsection{Decay Time and Associated Viscous Time}\label{sec:Disc_viscous}

In Section \ref{sec:decay_rise} the exponential decay rate of the near-IR brightness is measured as $0.74\,$yr for the most recent decline, using the Liverpool Telescope data set (Figure \ref{fig:LCs_850_LH}). This decay rate fits very well the previous decays after burst (Figure \ref{fig:LC_slopes}). Thus, it appears that the timescale has a physical origin which we associate with a viscous spreading time in the accretion disk surrounding the protostar, $\tau_{\nu_0}$. To within order unity  \citep{hartmann98},
\begin{equation}
    \tau_{\nu_0} = \frac{R_0^2}{3\,\nu_0},
\end{equation}
where
\begin{equation}
    \nu_0 = \alpha\, c_s\, H_0,
\end{equation}
and $R_0$ is the characteristic size of the accreting disk, $H_0$ is the scale height of the disk at $R_0$, $c_s$ is the sound speed in the disk, and $\alpha < 1$ is a scaling parameter for the viscosity \citep[see, e.g.][]{shakura73,matsuyama2003}.
%(See {Matsuyama et al.\ 2003 and other references}).

Rearranging, we obtain an equation for the characteristic size of the accreting disk
\begin{equation}
    R_0 = 3\, \alpha\, c_s \left( \frac{H_0}{R_0} \right) \tau_{\nu_0}.
\end{equation}

After substitution of fiducial values, 
we find
\begin{equation}
    R_0 = 0.15\, \alpha \left( \frac{c_s}{3\,{\rm km\,s}^{-1}} \right) \left( \frac{H_0/R_0}{0.1} \right)\, \left(\frac{\tau_{\nu_0}}{0.74\,{\rm yr}}\right){\rm AU}.
\end{equation}

Thus, it is possible to associate the decay time with the viscous time in the very inner region of the circumstellar disk, where enhanced accretion is taking place. This requires, however, both a very small inner accretion disk, $R_0 \sim 0.05$\,AU ($10 \,$R$_\odot$), {\it and} a very high viscosity parameter, $\alpha \sim 0.3$. We note that this $\alpha$ value is higher than the typical value accepted for FU Ori, $\sim 0.2-0.02$ \citep{zhu07}.
%The typical constraint of

\subsection{Mass Accretion During Burst}\label{sec:Mass_acc}

The amount of material accreted onto the protostar during a single burst can be determined directly from the bolometric light curve, under the assumption that the observed brightness is due to released accretion energy 
\citep{hartmann09book}.
Taking a mass of EC 53 of  $M_* \sim 0.3~{\rm M}_\odot$ \citep{lee20} and assuming a radius of $R_* \sim 2 {\rm R}_\odot$, we estimate a mass accretion rate of
\begin{equation}\label{eqn:ldot}
\dot M \simeq 
\left( \frac{4.2}{\eta} \right) 
\left( \frac{ L_{\rm acc}}{{\rm L}_\odot} \right)
\left( \frac{R_*}{{2\,\rm R}_\odot} \right)
\left( \frac{0.3\, {\rm M}_\odot}{M_*} \right)\,
\times 10^{-7}\ {\rm M}_{\odot}\, {\rm yr}^{-1},
\end{equation}
where  $\eta \leq 1$ captures the uncertainty in the fraction of energy radiated away during accretion. It is anticipated that $\eta \sim 1$ (see \citealt{hartmann11} for discussion).

The observed exponential decay rate of the near-IR brightness suggests a similar relationship of luminosity versus time,
\begin{equation}
    L_{\rm acc} = L_{\rm pk} \exp(-t/\tau_{\nu_0}),
\end{equation}
with $L_{\rm pk} \sim 20\, ~{\rm L}_{\odot}$ \citep{baek20}.
Substitution into Eqn.\ \ref{eqn:ldot} and integration over the length of the observed decay, $t_{\rm b} \approx 1.5\ $yr, we find
\begin{align}
    M_{\rm b} &= \int_{0}^{t_{\rm b}} \dot M dt \\
    &= \left( \frac{8.4}{\eta} \right) \tau_{\nu_0} \left( 1 - \exp(-t_{\rm b}/\tau_{\nu_0}) \right) \, \times 10^{-6} \ {\rm M}_{\odot} {\rm yr}^{-1},\\
    &= \left( \frac{5.4}{\eta} \right) \times 10^{-6} \ {\rm M}_{\odot}.
\end{align}

 With $\eta \sim 1$, the maximum accretion rate, $\dot  M_{\rm pk} \sim 8.4 \times 10^{-6}\ {\rm M}_{\odot}\, {\rm yr}^{-1}$, significantly larger than the values typically measured for Class I sources from direct accretion diagnostics \citep[e.g.][]{white04,salyk13} or from bolometric luminosities \citep{dunham10,fischer17} and even greater than the expected steady-state accretion value for a young, Class 0/I, protostar forming over $\sim 0.5\ $Myr.

\subsection{Inner Disk Mass}\label{sec:Disc_mass}

In the preceding sections we have assumed that a viscous inner accretion disk is responsible for the accretion of material onto the protostar. Thus, applying $\alpha$-disk constraints \citep[e.g.][]{hartmann98}, we estimate the required surface density, $\Sigma_0 = \Sigma(R_0)$, in the inner accreting disk:
\begin{align}\label{eqn:mdot}
    \dot M &= 3\, \pi\, \alpha\, c_s\, H_0\, \Sigma_0, \\
    &= \frac{\pi\, R_0^2\, \Sigma_0}{\tau_{\nu_0}}.
\end{align}
Rearranging,
\begin{align}\label{eqn:sigma}
    \Sigma_0 &= \frac{\dot M\, \tau_{\nu_0}}{\pi\, R_0^2}.
\end{align}
Furthermore, assuming steady-state $\alpha$-disk conditions yields $\Sigma(R) = \Sigma_0\,(R/R_0)^{-1}$ and therefore
\begin{align}
    M_d(R_0) &= 2\, \pi \, R_0^2\, \Sigma_0\,\\
    &= 2 \left( \dot M \, \tau_{\nu_0} \right),\\
    &= \left( \frac{1.2}{\eta} \right) \times 10^{-5}\ {\rm M}_\odot.
\end{align}

Thus, the mass of the inner accretion disk is only 2.3 times larger than the total amount of material accreted during a single episodic event suggesting that an outer disk reservoir is {\it required} to replenish the inner accretion disk.

\subsection{Outer Disk Conditions}

While the inner disk is observed to accrete material onto the protostar episodically, we have shown in the preceding section that there must be replenishment from the outer disk in order for the bursts to recur. If we assume that accretion through the outer disk is steady, then
\begin{align}
    \dot M_d &= \frac{M_{\rm b}}{t_{\rm b}},\\
    &= \left( \frac{3.6}{\eta} \right) \times 10^{-6}\  {\rm M}_\odot\,{\rm yr}^{-1}.
\end{align}
 Thus, even our assumed steady-state mass accretion rate in the outer disk reflects better anticipated Class 0 conditions versus Class I expectations.

Furthermore, the outer disk mass is estimated as $M_d \sim 0.07\ {\rm M}_\odot$ by \citep{lee20}, assuming standard conversions between millimeter continuum emission and mass. If the outer disk 
%is also assumed to be an $\alpha$-disk, such that 
 has a surface density $\Sigma \propto R^{-3/2}$, then 
\begin{align}
    M_d(R) = 0.07 \left( \frac{R}{R_d} \right)^{1/2} {\rm M}_\odot.
\end{align}
Starting with a large disk, $R_d \sim 100\,$AU \citep{lee20}, we extrapolate this outer disk mass to the scale of the inner accretion disk where $R_0 \sim 0.05\,$AU, 
finding $M_d(R_0) \sim 1.6 \times 10^{-3}\ {\rm M}_\odot$. This result is approximately 130 times {\it larger} than the above determined mass of the inner accretion disk (Section \ref{sec:Disc_mass}).

We thus conclude that while the outer disk has plenty of mass available near its inner edge, it is {\it unable} to smoothly supply the inner disk with material.  An accretion blockage occurs somewhere beyond the inner disk, $R > R_0 \sim 0.05\,$AU, where the inward accreting material from the outer disk is held up.  The blockage is only temporary, however, and after $\approx 1.5\ $yr the stored material is released and quickly drains onto and through the inner disk.

%An additional consequence of this model is that the $\alpha$ value required for the outer disk  is significantly lower than for the inner disk. If both the inner and outer disks have $\Sigma \propto  R^{-1}$, then the inner disk surface density normalization is fifteen times smaller. {\dij Assuming that the quantity $c_s\,H/R$ does not vary significantly} between the inner and outer disk then the ratio of $\alpha$ is inversely proportional to the surface density scale factor, implying that in the outer disk $\alpha \sim 0.02$ (see Eqn.\ \ref{eqn:mdot}).

 An additional consequence of this model is that the effective $\alpha$ value for the inner part of the outer disk is significantly lower than for the inner disk. Assuming that the quantity $c_s\,H$ does not vary significantly across the inner and outer disk  boundary then the ratio of $\alpha$ is inversely proportional to the local surface density scale factor (see Eqn. \ref{eqn:mdot}), implying that in the outer disk $\alpha \sim 0.002$.

\subsection{Replenishment Time and the Extinction Event}

\begin{figure*}[htp]
	\centering
	{\includegraphics[trim={0cm 0.0cm 0.0cm 0.0cm}, clip,width=175mm]{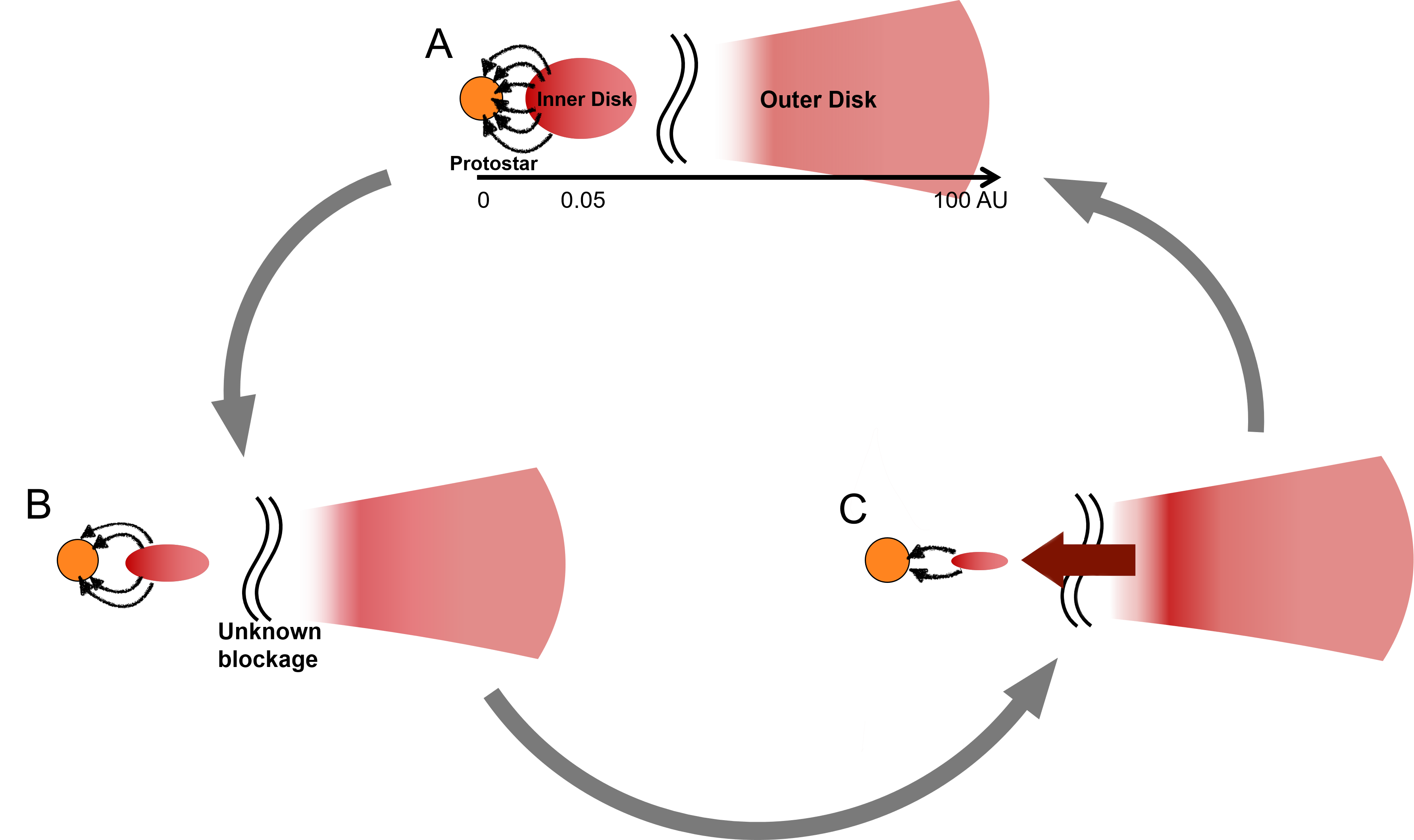}}
	\caption{Schematic view of the EC 53 system. A: The inner disk, at its fullest and thickest (Phase$=$0.5), begins accreting onto the central protostar. The protostar and inner disk brighten rapidly as the viscous inner disk spreads (Phases 0.5 -- 0.7). Nevertheless, the source appears reddest due to extinction produced by the puffed disk. B: Losing mass to the central protostar, the inner disk becomes thinner, the extinction lessens, the observed emission becomes less red, and the accretion rate declines (Phases 0.7 -- 1 \& 0 -- 0.3). During this stage  material from the outer disk is blocked from replenishing the inner disk. C:  Held back material from the outer disk quickly falls onto the inner disk during this short stage (Phases 0.3 -- 0.5), priming the system for the next cycle.}
\label{fig:Scheme}
\end{figure*}

From the near-IR lightcurve we find an exceptionally fast exponential rise time to the peak brightness: $t_r \approx 0.1\ $yr. Preceding the near-IR rise, however, we observe enhanced reddening of the source, which we associate with a strong extinction event (Figure \ref{fig:PD_CC}). The extinction begins around phase 0.25 and becomes extreme near phase 0.5. The full reddening event persists for 
%If we consider the entire reddening event we find a timescale of 
$\approx 0.4\ $yr, although the most extreme ($A_V\sim3$) reddening lasts for only $\approx 0.1$ yr.

%whereas if we only consider the extreme reddening we reproduce the lightcurve brightening time $t_r \sim 0.1\ $yr. 

If the material partaking in each individual burst is held up in the outer disk and then released, rapidly falling to the inner disk, then the fastest timescale available would be the dynamical time.  For a $0.3\  {\rm M}_\odot$ protostar, a $0.1\ $yr infall timescale corresponds to a distance of $R \sim 0.4\ $AU. This scenario requires both a mechanism for holding back the outer disk accretion at $\sim 0.4\ $AU, and a release mechanism that streams the material {\it directly} into the inner disk. Despite these complexities, an intriguing aspect of this scenario is that the burst periodicity, $t_{\rm b} \approx 1.5\ $yr, is only a few times the local Keplerian orbital time in the disk where the material is required to be held up (freefall time is $=\, 0.18\, \times$ the orbital time).

Alternatively, the viscous and accreting outer disk might be efficient to well inside $0.4\,$AU but become hung up closer to the measured size of the inner disk, $0.05\,$AU, with the effective viscosity at this boundary  decreasing steeply such that $\alpha << 1$. In this scenario the increased mass building up near the inner disk boundary takes place over many local Keplerian revolutions, eventually reaching a tipping point during which the effective viscosity rises sharply such that {$\alpha \sim 0.3$}.  We note that the physical plausibility of this scenario requires theoretical validation as the effective viscosity has to change dramatically over very short timescales.

The physical cause for the cycles of bursts and blockages is uncertain. Explanations for accretion bursts include magnetospheric instabilities in the inner disk \citep[e.g.][]{dangelo10}, gravitational instabilities near the star \citep[e.g.][]{bae14}, and interactions between the disk and a close companion \citep[e.g.][]{bonnell92,nayakshinlodato12,munoz16}.

The disk fragmentation scenario \citep[e.g.][]{vorobyovbasu10a} is especially intriguing for EC 53 given the estimated disk mass to protostar mass ratio of $\sim 0.25$.  This scenario, however, places the fundamental instabilities in the outer disk and would probably require longer timescales than measured for EC 53, although the migration of clumped material from the outer disk might be responsible for triggering an instability within the inner disk \citep{nayakshinlodato12}
.  

Of all the possible explanations, the planet or companion hypothesis suggested by \citet{hodapp12} and \citet{yoo17} is the only one that would specifically invoke quasi-periodic behavior.

Several other outbursts found in the literature have been analysed for their triggering mechanisms. The mid-IR outburst of Gaia 17bpi preceded the optical outburst, implying an outside-in event where the disk heated prior to accretion onto the central star \citep{hillenbrand18}. The outburst event of the Class I object Gaia 19ajj showed extinction from color variation in the mid-IR \citep{hillenbrand19}.  FU Ori itself may have a constant replenishment of its inner disk \citep{liu19_fuori}, leading to a smooth lightcurve for almost a century.
Dwarf novae, which undergo repetitive observable outbursts, allow for monitoring and provide considerable references \citep{osaki96}. For example, theoretical models have been used to investigate hysteresis observed in the time-dependent color-magnitude diagrams of some dwarf novae, attempting to fit simultaneously emission from the star, the accretion hot spot, and the disk with moderate success \citep{hameury20}. Unlike the hysteresis seen in the color-magnitude diagram for EC 53 (Figure \ref{fig:CMD_rel}), where the change in color with phase appears to be due to a changing extinction, for the dwarf novae the optical color change is interpreted as a variation in either the temperature of a component or a change in the relative brightness between components.

Previous studies typically focus on the burst, while in this paper we also demonstrate the relevance of the blockage.  In the schematic picture described here (Table \ref{tab:toy} and Figure \ref{fig:Scheme}), this blockage is understood as change in the disk viscosity from 
%$\alpha \sim 0.3$ in the most inner regions to $\alpha \sim 0.02$ in the outer regions.  
 high $\alpha$ in the most inner regions to low $\alpha$ in the outer regions. While our derived absolute value of $\alpha$ for the inner disk is significantly larger than normally assumed, its ratio against the estimated outer disk $\alpha$ indeed satisfies this requirement.

 An alternative possibility invokes changes in the mass-loss rate or pressure within the wind to destabilize disk accretion. \citet{lee15} demonstrated that wind pressure prevents the refilling of the disk when accretion is strongest for the early phase of the FUor burst of HBC 722.

Whatever the specific physical causes of the bursts and blockages, they must explain the rise and decay cycles, and the generic picture of the inner disk that we have derived from analysing those cycles.

\subsection{Toy Model to Explain the Observations}

Utilizing the observational evidence presented in the preceding section (Section \ref{sec:result}) we build here a simple toy model for the near-IR and sub-mm lightcurves of EC 53. We begin by assuming that the accretion-dominated bolometric luminosity lightcurve is composed of two coupled exponential functions:
\begin{align}
    \frac{L_{\rm acc(\phi)}}{L_{\rm pk}} = \left[ \exp( -\frac{\phi + \phi_0}{\Delta \phi_d}) +
    0.85\, \exp(-\frac{\phi + \phi_0 - 1}{\Delta \phi_r}) \right],
\end{align}
where $\phi$ is the phase (between 0 and 1), $\phi_0 = 0.35$ is an offset required to match the observed phase-folded diagram (Figure \ref{fig:PD_LC}), and the coefficients in the exponential functions reproduce a decay time of 0.74\ yr ($\Delta \phi_d = 0.52$) and a rise time of 0.2 yr ($\Delta \phi_r = 0.13$) when scaled to the 1.5 yr period of the system.  We further assume that the unattenuated near-IR luminosity scales with the accretion luminosity such that
\begin{align}
    m_{J_0}(\phi) &= -2.5\, \log(L_{\rm acc}(\phi))\,(1.28/1.3) + 13.9,\\
    m_{H_0}(\phi) &= -2.5\, \log(L_{\rm acc}(\phi)) + 11.8,\\
    m_{K_0}(\phi) &= -2.5\, \log(L_{\rm acc}(\phi))\,(0.76/0.7) + 10.0,
\end{align}
where the non-linear scaling is meant to create a slight color variation with changing source luminosity such that the unattenuated light from the system is somewhat bluer when least luminous.

\begin{figure}[htp]
   \centering
   {\includegraphics[trim={0.2cm 0.9cm 0.5cm 1.5cm},clip,width=0.98\columnwidth]{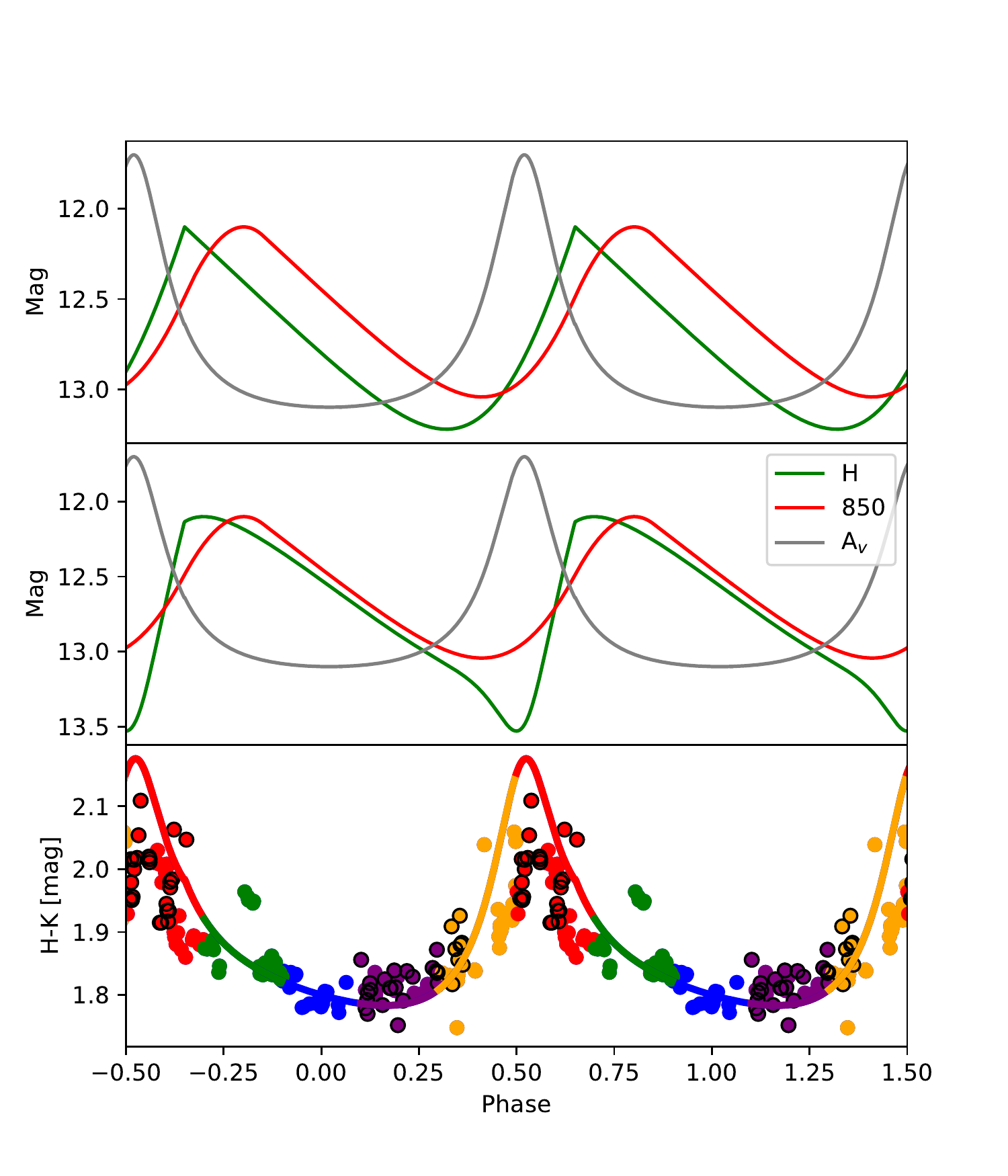}}
\caption{The light curves and color curve from the toy model. The top and middle panels show the $H$-band, sub-mm, and the extinction curve. The top panel shows the original light curve of $H$-band and the middle panel shows that after applying extinction. The bottom panel presents the color curve of $H-K$ from the toy model (line) with the observation (circles).}
\label{fig:Toy_F1}
\end{figure}

The sub-mm brightness is dominated by emission from dust in the extended obscuring envelope surrounding EC 53, which has a delayed reaction of up to a couple of months to changes in the accretion luminosity due to the light crossing time \citep{Johnstone13} for the envelope size of 10,000 AU \citep{baek20}. To simplify the computation for this model, we calculate the sub-mm flux by averaging the dust temperature response over $\delta \phi = 0.2$ and delayed by $\phi_s = 0.1$ ($0.15\,$yr). The top panel of Figure \ref{fig:Toy_F1} shows the resultant magnitudes light curve at $H$-band and in the sub-mm (after re-scaling). 
%It is  from the figure that the sub-mm light curve lags the unattenuated near-IR light curve at all phases.
 Due to the smoothing and delay, the sub-mm light curve extrema lag those in the unattenuated near-IR light curve at all phases.

\begin{figure}[htp]
   \centering
   {\includegraphics[trim={0cm 3.5cm 0.5cm 4.0cm},clip,width=0.98\columnwidth]{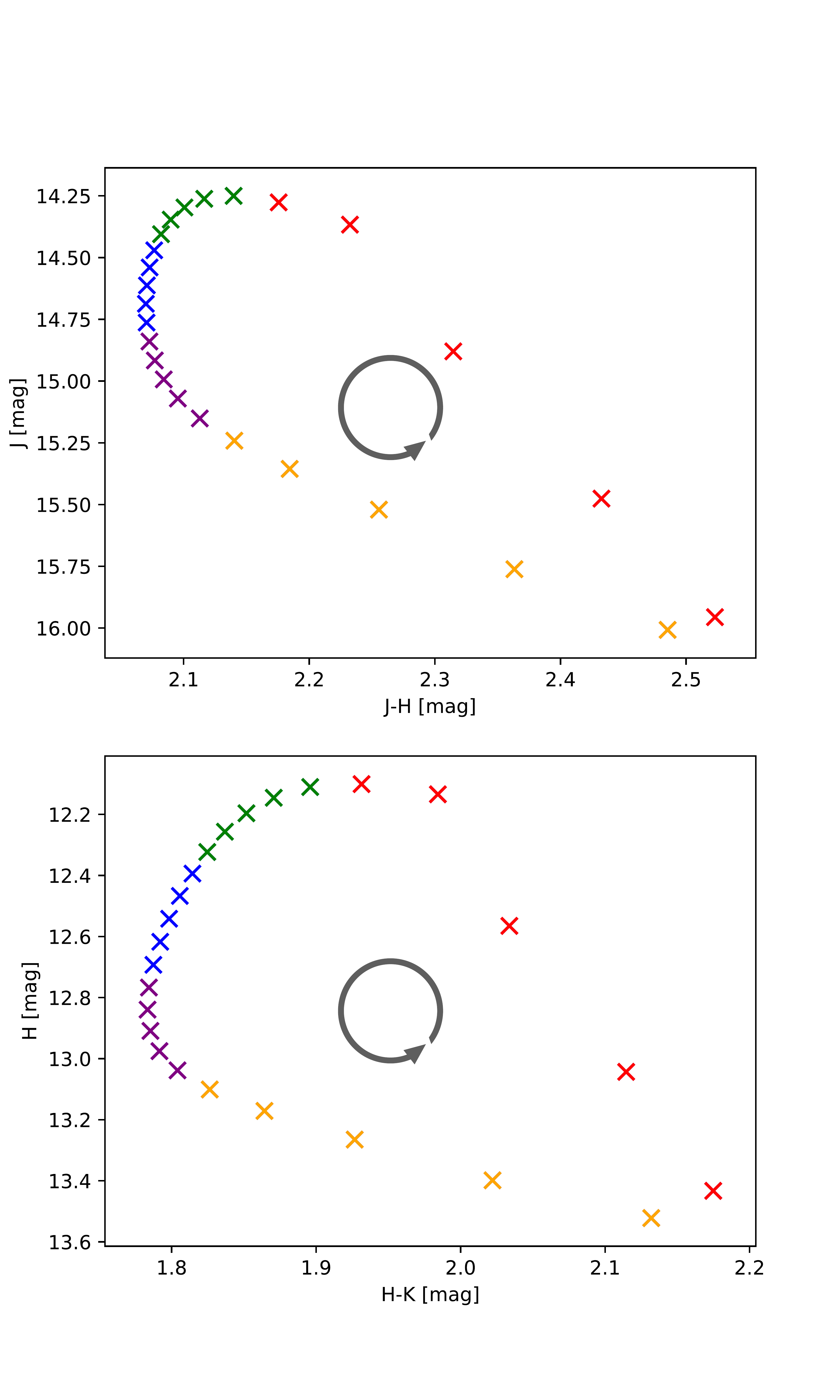}}
\caption{The Color-magnitude diagram obtained from the toy model. The color of the markers is tinted same as Figure \ref{fig:Toy_F1}}
\label{fig:Toy_F2}
\end{figure}

The observational data also reveals a strong reddening event just {\it after} the sub-mm emission begins to rise (Figure \ref{fig:PD_CC}), which we associate with enhanced extinction. For the toy model we represent the extinction event as a localized event, using the following equation:
\begin{align}
    A_J = \frac{1.5\,\gamma^2}{(\phi - \phi_A)^2 + \gamma^2},
\end{align}
where the maximum extinction takes place at $\phi_A = 0.52$ and $\gamma = 0.1$.

The effect of the extinction event on the observed near-IR magnitudes is thus
\begin{align}
    m_J &= m_{J_0} + A_J,\\
    m_H &= m_{H_0} + (0.155/0.265)\,A_J,\\
    m_K &= m_{K_0} + (0.090/0.265)\, A_J,
\end{align}
where the coefficients in front of the exponential reproduce the expected reddening law due to extinction \citep{cohen81}. 

The middle panel of Figure \ref{fig:Toy_F1} shows the resulting light curves. 
%The sub-mm rise is now {\it advanced} with respect to $H$-band, while the peak remains somewhat {\it delayed}, in agreement with the observations.
The sub-mm rise is now {\it advanced} with respect to $H$-band, entirely due to the addition of the near-IR extinction event. The sub-mm peak remains somewhat {\it delayed} due to the imposed propagation lag through the extended envelope. These features in the model light curves are now in agreement with the observations.
Furthermore the $H$-band rise has become significantly steeper than the sub-mm rise, which occurs at the same time as the extinction event. Compared with the observations, the most significant departure in the light curve produced by the toy model is the enhanced dip at the end of the H-band decay. In the observations the decay appears smoother, suggesting that the functional form for the extinction event is not quite appropriate. The bottom panel of Figure \ref{fig:Toy_F1} shows the $H-K$ color for the toy model, again reasonably reproducing the observations.

The toy model  color-magnitude diagrams are reproduced in Figure \ref{fig:Toy_F2}. In both panels we include measurements with a two week cadence in order to show the time spent in each part of the cycle. The overall appearance of the panels match the observations, despite significant differences in the details. 

%
%For comparison, the theoretical model for the dwarf novae outbursts using the disk parameter space did not actually reproduce the whole route on the color-magnitude diagrams \citep{hameury20}. The consideration of varying extinction caused by infalling mass seems to be necessary. 
%

%\begin{figure}
%   \centering
%   {\includegraphics[clip,width=1.05\columnwidth]{figures/Figure-XX.pdf}}
%\caption{To be completed}
%\label{fig:XX}
%\end{figure}

\section{Conclusions}\label{sec:Concl}

This paper summarizes our understanding of the eruptions and fades of EC 53.  This cycle faithfully repeats every $\approx 530$ days, in contrast to the apparent stochasticity of explosions in other eruptive YSOs.  This predictability has motivated us to intensely monitor EC 53 in the sub-mm and near-IR, supplemented by high-resolution ALMA imaging and high-resolution near-IR spectroscopy as well as radiative transfer modeling.

In our schematic interpretation of EC 53, steady accretion from the outer disk to the star is not possible.  Thus, mass builds up somewhere within the inner 0.4 AU of the disk. After about 1.5\,yr the blockage becomes unstable, draining into the inner disk and onto the star. 
This scenario is similar to that described by \citet{banzatti15} for the accumulation and release of mass in the EX Lup disk, as measured from H$_2$O, CO, and OH line emission \citep[see also][]{goto11}.   For EC 53, the total mass accreted during a burst is $\sim5\times 10^{-6}$ M$_\odot$, about half of the total mass of the inner disk.  Furthermore, the rise and decay e-folding timescales are always the same, indicating consistent physical explanations between epochs, despite some stochasticity in timing and amount of material accreted.

%The buildup of material is consistent with the increased  extinction, indicating that the disk scale height grows with mass.  
 The buildup of material would lead to a larger geometric height of the disk, consistent with the increased extinction.
This explanation is similar to occasional associations between increased accretion and extinction sometimes seen in FUors \citep[e.g.][] {kopatskaya13,hackstein15,hillenbrand18,hillenbrand19}. 
%and in smaller bursts (reference, not sure -- maybe Espaillat?).
For EC 53, the extinction increases sharply around the peak luminosity, and then disappears quickly as most of the material is drained, leading to a geometrically smaller disk. 

Our schematic picture places strong, generalized, constraints on the physical mechanisms that cause the cycles of eruption and decay for EC 53. The inner disk must be highly viscous to facilitate such large accretion rates to drain most of the inner disk.  The inner disk must be replenished sufficiently to trigger a burst 525 days later, from a region in the disk with a lower viscocity.  We hope that this schematic picture motivates investigation into the companion and/or instability physics that may cause repeated eruption and decays of young stellar objects.

%since our initial discovery of sub-mm quasi-periodicity by \citet{yoo17}, technically a rediscovery of near-IR periodicity found by \citet{hodapp99} and \citet{hodapp12}.  

%\citet{banzatti15}

%\citep{} -->  (Lee et al., 2019)
%\citet{} -->  Lee et al. (2019)
%\citealt{} --> Lee et al. 2019

\section*{Acknowledgement}
We thank the anonymous referee for thoughtful comments that helped to improve this paper.
This work was supported by the Basic Science Research Program through the National Research Foundation of Korea (grant No. NRF-2018R1A2B6003423) and the Korea Astronomy and Space Science Institute under the R\&D program supervised by the Ministry of Science, ICT and Future Planning. 
D.J. is supported by the National Research Council of Canada and by an NSERC Discovery Grant. T.N. and C.C.P. were supported by a Leverhulme project grant.   G.J.H. is supported by general grant 11773002 awarded by the National Science Foundation of China.

UKIRT is owned by the University of Hawaii (UH) and operated by the UH Institute for Astronomy. When (some of) the data reported here were acquired, UKIRT was supported by NASA and operated under an agreement among the UH, the University of Arizona, and Lockheed Martin Advanced Technology Center. The operations were enabled through the cooperation of the East Asian Observatory.  The authors wish to recognize and acknowledge the very significant cultural role and reverence that the summit of Maunakea has always had within the indigenous Hawaiian community.  We are most fortunate to have the opportunity to conduct observations from this mountain.

We are grateful to Robert Smith for useful information on the Liverpool Telescope $H$-band system response.
The Liverpool Telescope is operated on the island of La Palma by Liverpool John Moores University in the Spanish Observatorio del Roque de los Muchachos of the Instituto de Astrofisica de Canarias with financial support from the UK Science and Technology Facilities Council.

Infrared photometric data on EC~53 were obtained at the IRIS telescope of the Universit\"atssternwarte Bochum on Cerro Armazones, which is operated under a cooperative agreement between the "Astronomisches Institut, Ruhr Universit\"at Bochum", Germany and the Institute for Astronomy, University of Hawaii, USA.

Construction of the IRIS infrared camera was supported by the National Science Foundation under grant AST07-04954.

\bibliographystyle{aasjournal}
\bibliography{paper}

\appendix

\setcounter{figure}{0} \renewcommand{\thefigure}{A.\arabic{figure}}

\section{Observational Details}\label{sec:A_Obs}

\subsection{SCUBA2}

The data reduction was performed using the iterative map-making software, {\sc{makemap}} \citep[see ][ for details]{chapin13}, which is part of {\sc{starlink}}'s \citep{currie14} Submillimetre User Reduction Facility ({\sc{smurf}}) package \citep{jenness13}. Post-processing was applied to each image in order to align them more precisely with one another and to apply a relative flux calibration. The alignment and relative flux calibration were carried out by analysing the positions and fluxes of bright, compact, point-like objects across the field and making relative adjustments based on measurements of non-variable sources \citep{mairs17b}. Finally, to mitigate pixel to pixel noise, a smoothing was performed using Gaussians with Full Width at Half Maxima of 4$\arcsec$ and 6$\arcsec$ for the 450 $\micron$ and 850 $\micron$ images, respectively. All specifics on the data reduction, image alignment, and flux calibration techniques are described in detail by \cite{mairs17b} (reduction \textit{R3}).

\subsection{UKIRT}\label{sec:A_UKIRT}
The observations were performed by dithering the object to five positions separated by a few arcseconds, and a 2 $\times$ 2 microstep for each dither. This resulted in a pixel scale of 0.2$\arcsec$ per pixel for the final mosaics.  For all epochs, observations in each filter was obtained with two different individual exposure times per frame: 2-sec, 2-coadd and 10-sec, 1-coadd in $J$, 1-sec, 2-coadd and 10-sec, 1-coadd in $H$, and 1-sec, 2-coadd and 5-sec, 2-coadd in $K$ respectively.

The sky conditions were not always clear, many of the observations were obtained in the presence of thin or slightly thicker cloud conditions. Preliminary reduction of the data was performed by the Cambridge Astronomical Survey Unit (CASU). The details of the data processing, photometric system and calibration are given in \cite{irwin04}, \cite{hewett06}, and \cite{hodgkin09}, respectively. Further reduction and aperture photometry were performed using the {\sc STARLINK} packages {\sc KAPPA} and {\sc PHOTOM}. As the sky conditions were not always photometric, relative photometry of EC 53 was obtained using 40 isolated point sources in the field of the same array in which it was located. Care was taken to avoid comparison stars which exhibited variability during the period of our monitoring. Aperture photometry was done in four different apertures, with radii 1.29$\arcsec$,1.5$\arcsec$, 3.68$\arcsec$ and 4$\arcsec$ respectively. The magnitudes were tied to the $JHK$ magnitiudes of the comparison stars measured in the UKIDSS survey.  Our monitoring spanned over a period of four outburst cycles. The light curves with the 3.68$\arcsec$ aperture are presented in Figure \ref{fig:LC_all_mag}. The error limits given are the 1-sigma errors in the mean zero points of the comparison stars.

\subsection{Liverpool Telescope}\label{sec:A_LT}

Although nominally $H$-band observations, the red edge of the IO:I $H$-band system response is defined by the sensitivity cut-off of the Teledyne Hawaii-2RG HgCdTe Array.
There is no detailed response for the detector, but it is known that its quantum efficiency is only 50 percent at 1.72 $\mu$m{\footnote{http://telescope.livjm.ac.uk/TelInst/Inst/IOI/}}, compared with a system throughput of 50 percent of peak throughput at 1.785 $\mu$m for the UKIRT system \citep{hewett06}.
Given that the IO:I $H$-band has an approximately standard blue cut-off at 1.49 $\mu$m, this  implies the width of the IO:I $H$-band is roughly three-quarters of that for a standard $H$-band.

Each night of observations consisted of several images (usually 9) taken with the telescope moving 14 arcseconds in RA and/or declination between each.
These individual images were cleaned using the standard Liverpool Telescope IO:I pipeline described by \cite{barnsley16}.
We measured the positions of all the sources in the field using a mean image from one night, and then carried out aperture photometry at those fixed positions (i.e. ``forced photometry") after allowing for offsets and rotations between images.

Having extracted photometry for each star from each image it is standard practice to use an ensemble of non-variable comparison stars to normalize the flux from each image to allow for differences in seeing (which will change the fraction of light which lies outside the aperture) and transparency.
There are two issues which make this problematical for the current data. 
First the IO:I pipeline removes the night sky emission using a median of the other images of the target taken on that night (a  ``peers-only'' median).
Whilst the 14 arcsec throw is sufficient to ensure that a small nebula such as EC 53 is moved far enough to obtain a clean sky measurement, there is a much larger region of nebulosity in the field which suffers poor night sky removal, and therefore poor photometry for those stars.
Second, the field is relatively sparse as the extinction is high, and the field-of-view of IO:I relatively small (6.27 by 6.27 arcminutes), exacerbating further the lack of comparison stars.

We were, therefore, careful to examine the precision of the resulting photometry to characterize these effects.
We first used all the measurements for each star to derive a $\chi^2$ with respect to a constant using uncertainties derived from the scatter between sky pixels \citep[e.g.][]{naylor98}.
We found that we had to add a systematic uncertainty of 0.03 mags to ensure $\chi^2$ did not increase with the mean flux of the star.
This allowed us to remove stars with a reduced $\chi^2$ of greater than 10 from the normalization procedure on the assumption they were variables, or suffered from poor background removal.
We then combined the measurements of each star on each night to create a nightly mean, and rather than using the formal uncertainties, used the scatter between the measurements to create a standard error.
Examining the lightcurves of the remaining stars suggests that our photometry of EC 53 is limited to a precision of 0.04 mags at $H$=12 mag for a nightly mean, with the possibility of systematic changes at this level on timescale of months.

\setcounter{figure}{0} \renewcommand{\thefigure}{B.\arabic{figure}}

\section{Calculation of the Periodicity of EC 53}\label{sec:A_Per}
We present the Autocorrleation Function (ACF) and Lomb-Scargle Periodogram (LSP) results for EC 53, with and without the offset determined by the phase-folded diagram string length method in Section \ref{sec:PD_rst}. Given that the sub-mm observations span a shorter range in time than the near-IR observations, we calculate the near-IR ACF and LSP both using entire data set and using only those times in common (i.e.\ after 2016 Feb 3 in UT).

%\begin{figure}[htp]
%   \centering
%   \subfigure{\includegraphics[trim={0cm 1.4cm 0cm 2.5cm},clip,width=0.8\columnwidth, keepaspectratio]{figures/EC53_ACF_cut.pdf}}
%\caption{The ACFs of near-IR and 850 $\mu$m bands. Only with the data observed after JD = 2457421, The orange lines presents the peak location which corresponds to the period earned by ACF; 555 for near-IR bands, and 570 for 850 $\mu$m band.}
%\label{fig:ACF_cut}
%\end{figure}

\subsection{ACF Analysis}\label{sec:ACF_rst}

The ACF is a useful tool to determine the period of light curves regardless of their underlying shape. The ACF can be derived by correlating a set of time-series data with itself, shifted by a regular time interval. We adopt equation (1) of \citet{mcquillan13} for calculating the ACF of the light curves. In the case of well-sampled periodic data, a clear periodic pattern appears in the ACFs. The position of the second local maximum in the ACF represents the period of variation. Although the ACF requires a set of regularly sampled data, it provides robust and clear results for periodic signals.

\begin{figure}[htp]
   \centering
   \subfigure{\includegraphics[trim={0cm 1.4cm 0cm 2.5cm},clip,width=0.95\columnwidth, keepaspectratio]{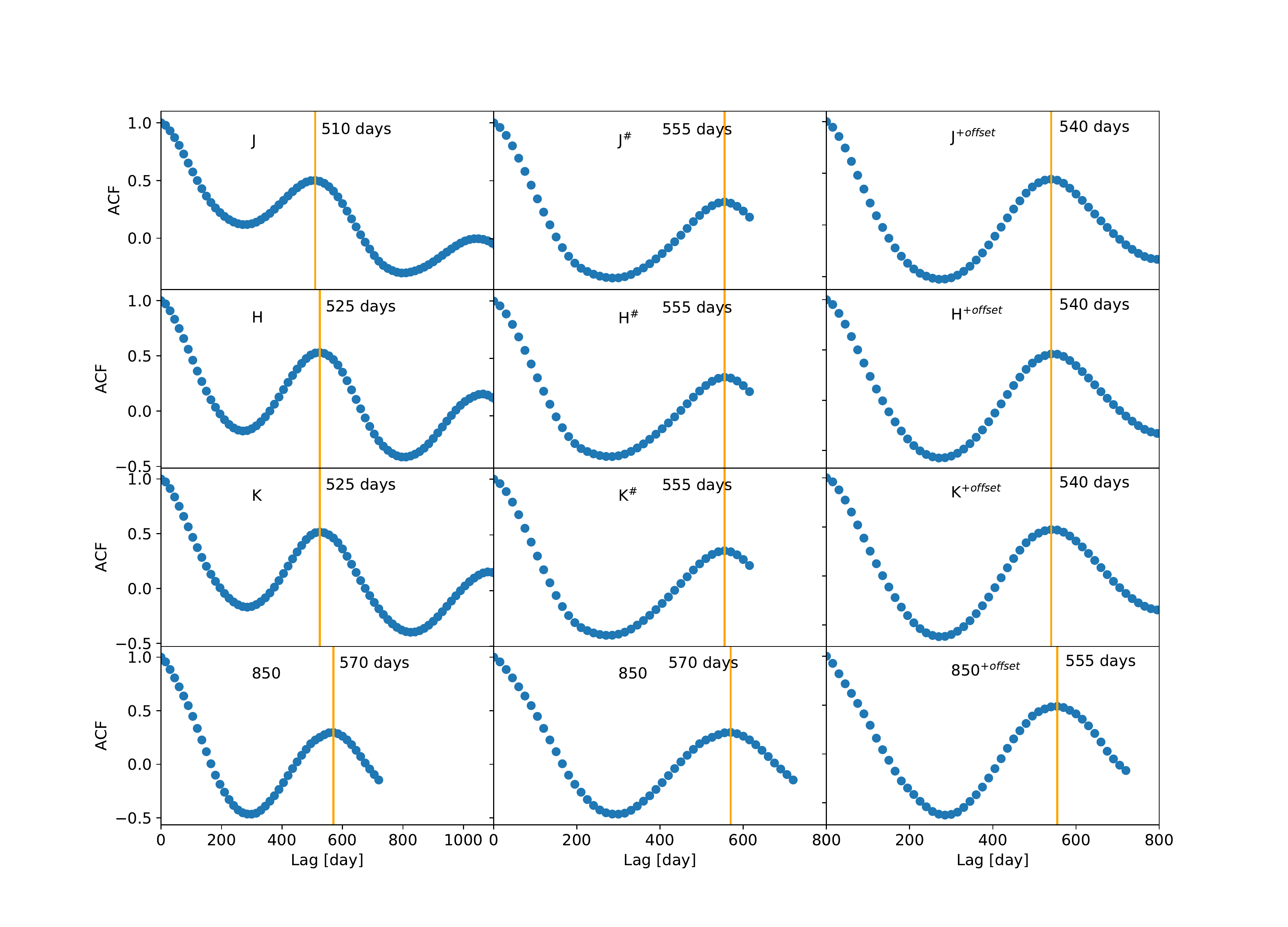}}
\caption{The ACFs of $J$, $H$, $K$ and 850 $\mu$m passbands, from all data set we have (left), in a common time baseline with 850 $\mu$m (middle), and after the offset subtraction for the first period (right). The orange lines show the peak location which corresponds to the period earned by ACF. The peak locations are 510, 525, 525, and 570 days for each passband when we used the whole data set. Using common time baseline yields 555 days at near-IR bands, and 570 days at 850 $\mu$m band. The 540, 540, 540, and 555 days for each passband after the offset subtraction.}
\label{fig:ACF_all}
\end{figure}

Before computing the ACF, the light curves are regularly resampled to have a constant time interval of fifteen days, using a linear interpolation. 
We calculate the ACF for the light curves at near-IR and 850 $\mu$m in three different ways (Figure \ref{fig:ACF_all}): (1) the entire data set at each wavelength (left panels), (2) a common time baseline with 850 $\mu$m (middle panels), and (3) the entire data set at each wavelength after removal of the step-function determined by the string length method of Section \ref{sec:PD_rst} (right panels). 

We identify the position of the second peak in each ACF as the period of the light curve. The uncertainty of the period is estimated at a half of the sampling interval (7.5 days). Under the time window constraint but no offset  (Figure \ref{fig:ACF_all} middle panel), the ACF periods show reasonable agreement across wavelength, with periods of $\approx 550\ $d at all near-IR bands and $\approx 570\ $d in the sub-mm. Applying the offset from Section \ref{sec:PD_rst}, makes the agreement even better, 540 versus 555\ d, respectively (Figure \ref{fig:ACF_all} right panel). For the unmodified, full near-IR data set, we find the ACF-determined periods are significantly shorter, 510 to 525\ d (Figure \ref{fig:ACF_all} left panel). The results are presented in tabular form in Table \ref{tab:PD_rst}.

\subsection{Lomb-Scargle Periodogram Analysis}\label{sec:LS_rst}

%\begin{figure}[htp]
%	\centering
%	{\includegraphics[trim={0cm 1.7cm 0cm 2.5cm},clip,width=0.8\columnwidth]{figures/EC53_Periodograms_cut.pdf}}
%	\caption{Periodograms of light curves in four different passbands; J, H, K, and 850 $\mu$m. The common time baseline obtained from the 850 $\mu$m observation is used for the near-IR passbands. The peak positions around 543 days are marked with orange vertical lines; 553, 553, 553, and 589 days for J, H, K, and 850 $\mu$m bands, respectively. Gray solid horizontal line indicate the power where the corresponding FAP is 10$^{-3}$ $\%$.}
%\label{fig:PS_cut}
%\end{figure}

We also apply a Lomb-Scargle periodogram (also called a least-squares spectral analysis; \citealt{lomb76}; \citealt{scargle89}) technique to the observed light curves to determine their periods using the \textit{LombScargle} task in the \textit{timeseries} package of \textit{astropy} \citep{astropycollabo13}. The LSP is a powerful tool to detect periodic signals in discrete time-series data. The False Alarm Probability (FAP) is calculated as described in \citep{baluev08}. For a detailed description, we refer to \citet{vanderplas15} and \citet{vanderplas18}. The best fit periods along with the uncertainty in the period, determined through Gaussian fitting to each peak in the frequency domain, are tabulated in Table \ref{tab:PD_rst}.

\begin{figure}[htp]
	\centering
	{\includegraphics[trim={0cm 1.7cm 0cm 2.5cm},clip,width=0.95\columnwidth]{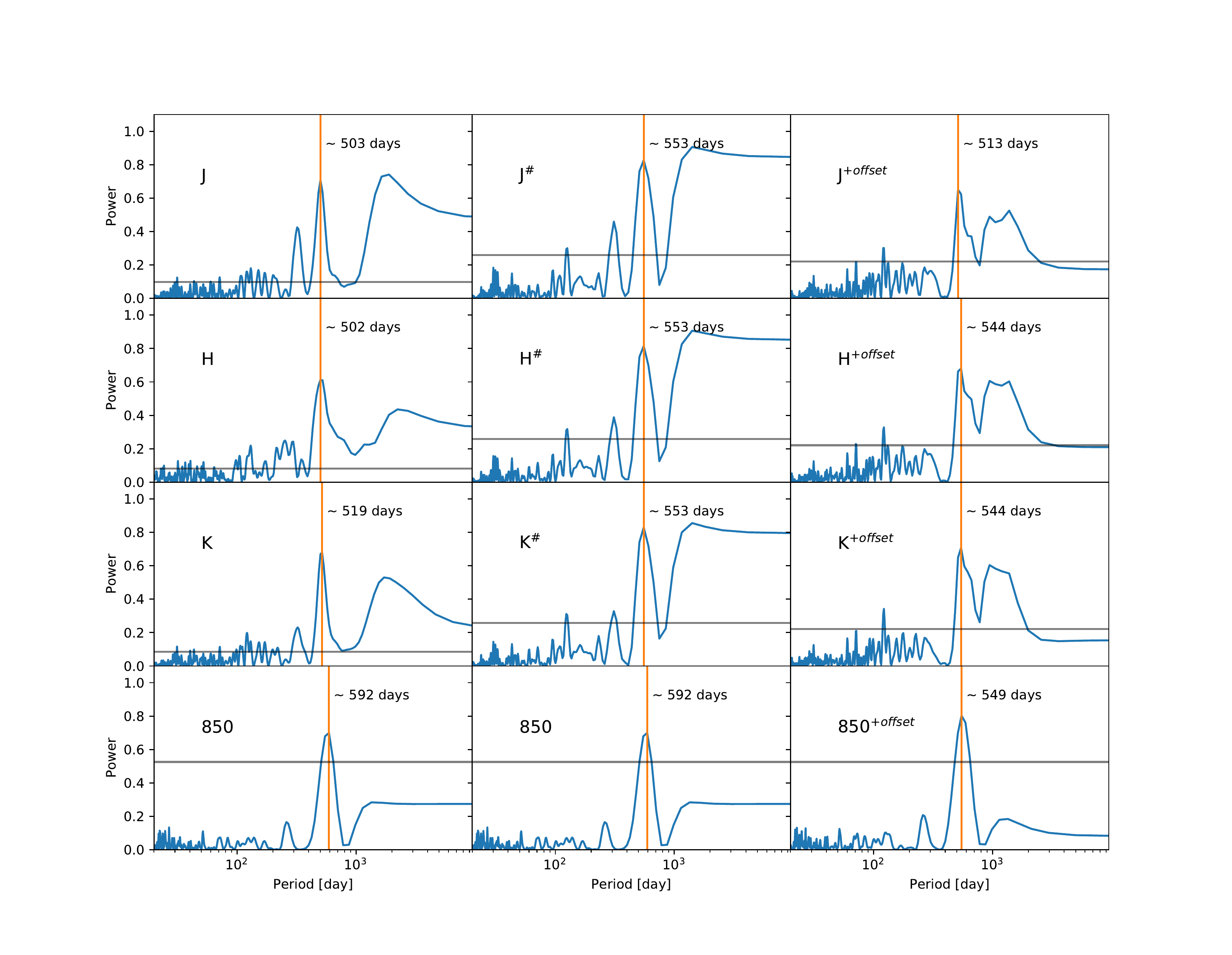}}
	\caption{Periodograms of light curves in four different passbands; J, H, K, and 850 $\mu$m from all data set we have (left), in a common time baseline with 850 $\mu$m (middle), and after the offset subtraction for the first period (right). The peak positions around 543 days are marked with orange vertical lines. The entire data set shows their peak at 503, 502, 519, and 592 days. The common time baseline with 850 $\mu$m shows 553, 553, 553, and 592 days. The step function applied lightcurves yield 513, 544, 544, and 549 days for each J, H, K, and 850 $\mu$m bands. The peaks appeared at $\approx$ 1700 days diminish after applying the step function. Gray solid horizontal line indicate the power where the corresponding FAP is 10$^{-3}$ $\%$.}
\label{fig:PS_all}
\end{figure}

The LSP-determined periods are very similar to the ACF periods in all cases. For the full, unmodified, near-IR data set an additional long period signal, $\approx 4\ $yr, is found, along with the $\approx 500\ $d period (Figure \ref{fig:PS_all} left panel). Common time baseline with 850 $\mu$m but without the offset, this additional long-period component endures (Figure \ref{fig:PS_all} middle panel). Applying the string length determined offset to the data prior to 2016 Sep 9 (Section \ref{sec:PD_rst}), however, removes the long-term component and brings the determined periods, $\approx$ 540 days, into agreement with the sub-mm,  the ACF, and the string length method of Section \ref{sec:PD_rst} (Figure \ref{fig:PS_all} right panel). We note that the FAP of the peaks obtained by LSP are well below the 10$^{-6}$ \% level.

\subsubsection{Periodogram Window Function}\label{sec:A_win}

\begin{figure}[htp]
   \centering
   \subfigure{\includegraphics[trim={0cm 0cm 0cm 0cm}, clip, width=0.48\columnwidth, keepaspectratio]{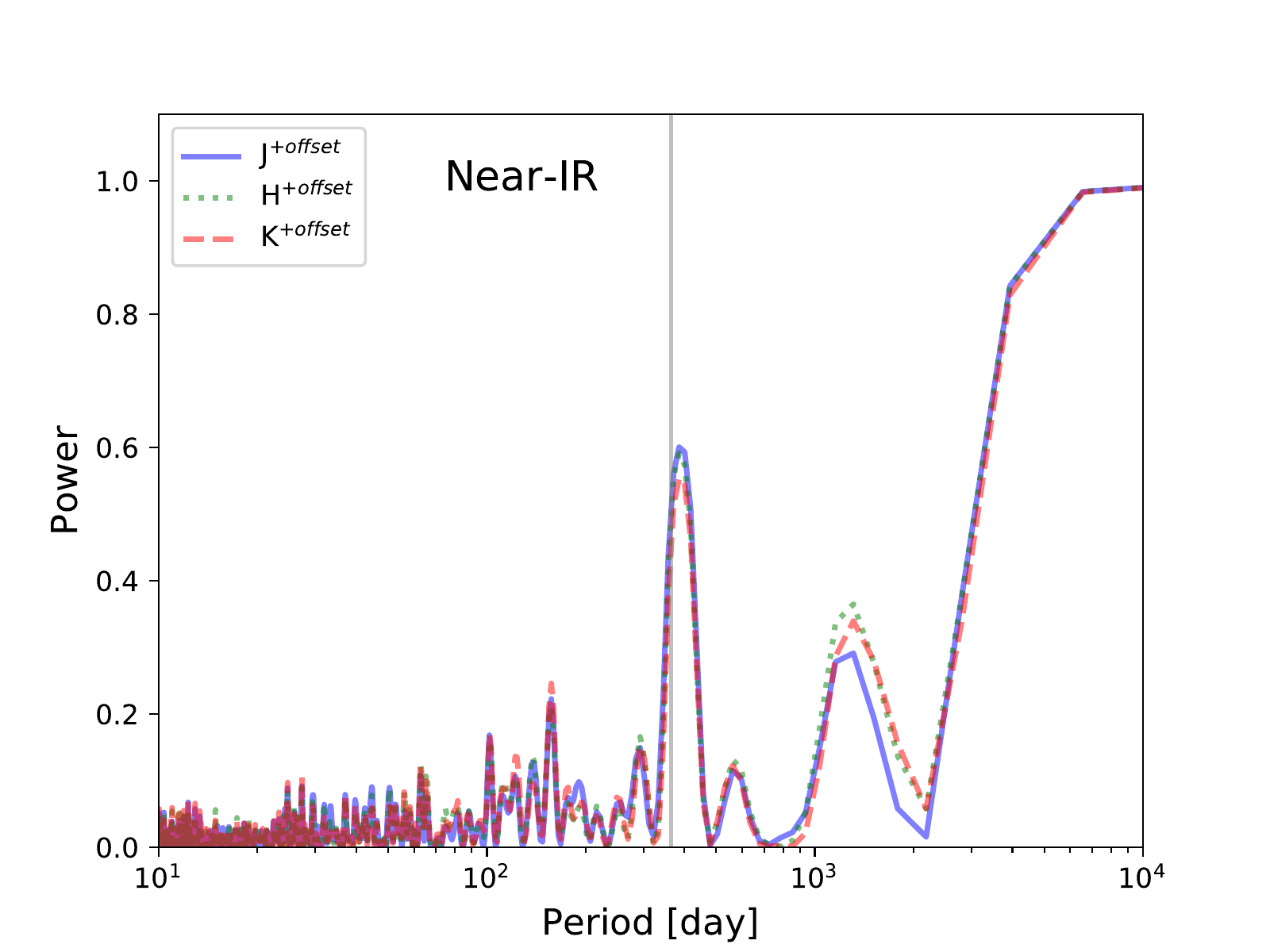}}
   \subfigure{\includegraphics[trim={0cm 0cm 0cm 0cm}, clip, width=0.48\columnwidth, keepaspectratio]{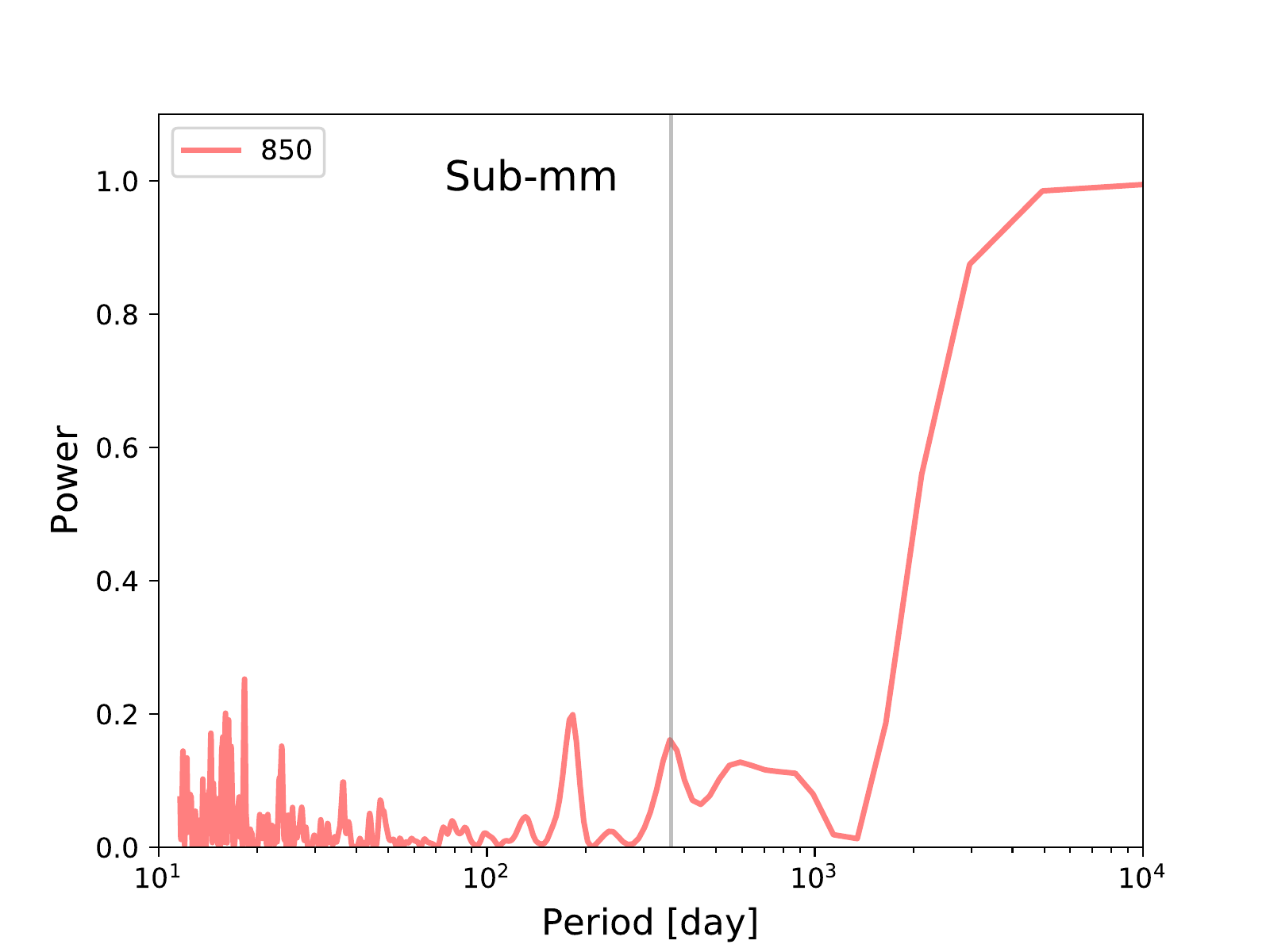}}
\caption{The window periodograms of JHK (left) and 450 $\&$ 850 $\mu$m (right) bands. The gray solid line marks a year period.}
\label{fig:PS_windows}
\end{figure}

We obtained periodograms of the window function from the data set used for Appendix \ref{sec:LS_rst} following Section 7.3.1 of \citet{vanderplas18}.  As shown in Figure \ref{fig:PS_windows}, there are notable peaks at $\approx$ 1\ yr and $\approx$ 3\ yr in the  window function periodograms of near-IR bands, reflecting the annual observing window for this source. For 850 $\mu$m, only minor peaks at a year, $\approx$ 200\,days and $\approx$ 15\,days appear. The 15\,days peak corresponds to our observational cadence. None of the LSP-determined periods for EC 53 appears to be influenced by the window function.

%These peaks seem to affect the near-IR periodogram results which gave shorter periods than 850 $\mu$m's result.

\begin{deluxetable*}{ccccccc}[hbp]
\tablecaption{The result of period determination \label{tab:PD_rst}}
%\tablecolumns{6}
%\tablenum{1}
%\tablewidth{0pt}
\tablehead{
\colhead{Bands} & \colhead{LSP} & \colhead{ACF} & \colhead{LSP$^{1}$} & \colhead{ACF$^{1}$} & \colhead{LSP$^{2}$} & \colhead{ACF$^{2}$}
}
\startdata
J & 503$_{-42}^{+49}$, 1894$_{-541}^{+1405}$ & 510$\pm$7.5 & 535$_{-72}^{+98}$ & 555$\pm$7.5 & 513$_{-42}^{+49}$ & 540$\pm$7.5 \\
H & 502$_{-62}^{+83}$, 1892$_{-417}^{+900}$ & 525$\pm$7.5 & 535$_{-70}^{+94}$ & 555$\pm$7.5 & 544$_{-65}^{+85}$ & 540$\pm$7.5 \\
K & 519$_{-41}^{+48}$, 1719$_{-438}^{+949}$ & 525$\pm$7.5 & 535$_{-67}^{+89}$ & 555$\pm$7.5 & 544$_{-62}^{+80}$ & 540$\pm$7.5
\\
850 $\mu$m & ... & ... & 592$_{-73}^{+98}$ & 570$\pm$7.5 & 549$_{-77}^{+105}$ & 555$\pm$7.5 
\\
\enddata
\tablenotetext{1}{Data points observed after JD $=$ 2457421 are used.}
\tablenotetext{2}{Data points obtained by UKIRT and the offset given before JD $=$ 2457640}
%\tablecomments{The.}
\end{deluxetable*}

%At K we get 508$_{-40}^{+46}$ days in $K$ band. The periodograms for near-infrared bands result in two peaks with the FAP less than 10$^{-3}$ \%; one peak is located around 4 years period, and the other peak appears around 510 days (orange solid lines in Figure \ref{fig:PS_all}). The rest other peaks are not thought to trace the nature of the system but could be produced due to the window function (see Appendix \ref{sec:A_win}) or harmonics of the dominant peaks.

%With longer time baseline, the periodograms of near-IR light curves gave more peak-like feature at the long-term component compare to ambiguous bump at sub-mm. This long term variation may cause the difference in the yielded period between our near-IR light curves and the K-band light curve in \citet{hodapp12}. For comparison we performed the ACF and periodogram analyses for near-IR light curves within the time baseline of 850 $\mu$m light curve (Table \ref{tab:PD_rst}). The periodogram results were $\sim$ 555 days which are more closer to that of 850 $\mu$m. The ACF shows the same value (570$\pm$7.5 days) in four wavelengths. This indicates that the regularity of the system is somewhat perturbed. We discussed about this perturbation in Section \ref{sec:DC_Non-r}. %The longer period derived from the 850 $\mu$m light curve might be caused by the combination of imcomplete observation coverage and non-sinusoidal functional variation.

\end{document}